\documentclass[twocolumn]{aastex631}

\usepackage{amsmath}
\usepackage{amssymb}
\usepackage{multirow}
\usepackage{bm}

\newcommand{\SO}{\operatorname{SO}}

\shorttitle{Wavelet Moments for Cosmological Parameter Estimation}
\shortauthors{Eickenberg et al.}

\graphicspath{{./}{figures/}}

\begin{document}

\title{Wavelet Moments for Cosmological Parameter Estimation}

\author{Michael Eickenberg}
\correspondingauthor{Michael Eickenberg}
\email{meickenberg@flatironinstitute.org}
\affiliation{Flatiron Institute Center for Computational Mathematics, 162 5th Ave, 3rd floor, New York, NY 10010, USA}

\author{Erwan Allys}
\affiliation{Laboratoire de Physique de l'École Normale Supérieure, ENS, Université PSL, CNRS, Sorbonne Université, Université Paris Cité, 75005 Paris, France}

\author{Azadeh Moradinezhad Dizgah}
\affiliation{D\'epartement de Physique Th\'eorique, Universit\'e de Gen\`eve, 24 quai Ernest Ansermet, 1211 Gen\`eve 4, Switzerland}

\author{Pablo Lemos}
\affiliation{Department of Physics and Astronomy, University of Sussex, Sussex House, Falmer, Brighton, BN1 9RH, UK}
\affiliation{Department of Physics and Astronomy, University College London, Gower Street, London, WC1E 6BT, UK}

\author{Elena Massara}
\affiliation{Waterloo Centre for Astrophysics, University of Waterloo, 200 University Ave W, Waterloo, ON N2L 3G1, Canada}
\affiliation{Department of Physics and Astronomy, University of Waterloo,
200 University Ave W, Waterloo, ON N2L 3G1, Canada}

\author{Muntazir Abidi}
\affiliation{D\'epartement de Physique Th\'eorique, Universit\'e de Gen\`eve, 24 quai Ernest Ansermet, 1211 Gen\`eve 4, Switzerland}

\author{ChangHoon Hahn}
\affiliation{Department of Astrophysical Sciences, Peyton Hall, 4 Ivy Lane, Princeton University, Princeton, NJ 08544}

\author{Sultan Hassan}
\affiliation{Flatiron Institute Center for Computational Astrophysics, 162 5th Ave 5th floor, New York, NY 10010, USA} \affiliation{Department of Physics \& Astronomy, University of the Western Cape, Cape Town 7535,
South Africa}

\author{Bruno R\'egaldo-Saint Blancard}
\affiliation{Flatiron Institute Center for Computational Mathematics, 162 5th Ave, 3rd floor, New York, NY 10010, USA}

\author{Shirley Ho}
\affiliation{Flatiron Institute Center for Computational Astrophysics, 162 5th Ave 5th floor, New York, NY 10010, USA}

\author{St\'ephane Mallat}
\affiliation{Coll\`ege de France, 11, place Marcelin Berthelot,
75231 Paris Cedex 05, France}
\affiliation{D\'epartement d'Informatique, École normale supérieure - PSL,
45 rue d’Ulm, F-75230 Paris cedex 05, France}
\affiliation{Flatiron Institute Center for Computational Mathematics, 162 5th Ave, 3rd floor, New York, NY 10010, USA}

\author{Joakim And\'en}
\affiliation{Department of Mathematics, KTH Royal Institute of Technology, SE-100 44 Stockholm, Sweden}
\affiliation{Flatiron Institute Center for Computational Mathematics, 162 5th Ave, 3rd floor, New York, NY 10010, USA}

\author{Francisco Villaescusa-Navarro}
\affiliation{Flatiron Institute Center for Computational Astrophysics, 162 5th Ave 5th floor, New York, NY 10010, USA}
\affiliation{Department of Astrophysical Sciences, Princeton University, Peyton Hall, Princeton NJ 08544, USA}


\begin{abstract}
Extracting non-Gaussian information from the non-linear regime of structure formation is key to fully exploiting the rich data from upcoming cosmological surveys probing the large-scale structure of the universe. However, due to theoretical and computational complexities, this remains one of the main challenges in analyzing observational data.  We present a set of summary statistics for cosmological matter fields based on 3D wavelets to tackle this challenge.
These statistics are computed as the spatial average of the complex modulus of the 3D wavelet transform raised to a power $q$ and are therefore known as invariant wavelet moments. 
The 3D wavelets are constructed to be radially band-limited and separable on a spherical polar grid and come in three types: isotropic, oriented, and harmonic.
In the Fisher forecast framework, we evaluate the performance of these summary statistics on matter fields from the Quijote suite, where they are shown to reach state-of-the-art parameter constraints on the base $\Lambda$CDM parameters, as well as the sum of neutrino masses.
We show that we can improve constraints by a factor 5 to 10 in all parameters with respect to the power spectrum baseline.
\end{abstract}


\section{Introduction} \label{sec:intro}

Cosmic large-scale structure (LSS) captures a wealth of information about the initial conditions, the composition, and the physical laws governing the evolution of the universe. Upcoming galaxy surveys, such as DESI \citep{Aghamousa:2016zmz}, Euclid  \citep{Amendola:2016saw}, SPHEREx \citep{Dore:2014cca}, and Rubin Observatory \citep{Abell:2009aa}, will map the LSS over very large cosmological volumes, at an unprecedented precision. These data offer an opportunity for shedding light on key open questions in modern cosmology, namely the origin of primordial fluctuations, the nature of dark matter and dark energy, and properties of neutrinos. Given the nonlinear nature of gravitational evolution forming the LSS, exploiting this rich data to its full potential relies on optimal extraction of non-Gaussian information beyond the linear scales. Focusing on the parameters of the $\Lambda$CDM model and the sum of the neutrino masses, we present in this paper a new set of wavelet-based statistics and assess their performance in constraining cosmological parameters on simulated data.

The measurements of two squared-mass differences between the three neutrino species by existing neutrino flavor oscillation experiments \citep{deSalas:2017kay,Esteban:2018azc} provide strong evidence for physics beyond the standard model, which predicts three massless neutrinos. These measurements have not determined the absolute mass of neutrinos and allow for two possible mass splittings -- one heavy and two light neutrinos (normal hierarchy), and one light and two heavy neutrinos (inverted hierarchy). Future oscillation experiments (e.g. \citealt{Hyper-Kamiokande:2018ofw,DUNE:2020jqi}) promise unambiguous determination of the neutrino mass hierarchy \citep{Patterson:2015xja}, while cosmological observations are expected to provide tight constraints on the absolute mass scale of neutrinos by measuring the sum of their masses (see \citealt{Lesgourgues:2018ncw} for an extensive review). Since the oscillation experiments put a bound on the minimum neutrino mass of $60 \ {\rm meV}$ in normal and $100\ {\rm meV}$ in inverted hierarchies, high-precision constraints on total neutrino masses from cosmology can indirectly constrain the mass hierarchy by excluding the inverted hierarchy. Observations of the LSS in the coming years provide the most promising means of reaching such a sensitivity, using the unique features they induce on the dark matter distribution and its biased tracers, such as galaxies: at linear order they do not cluster below their free-streaming scale, suppressing the growth of structure on small scales.

To date, most of information from cosmological surveys is obtained from measurements of 2-point statistics (Fourier or angular power spectra and configuration-space correlation functions). For a statistically homogeneous (i.e., stationary) Gaussian field, the power spectrum fully characterizes the statistical properties of the field and therefore encodes all the information about the parameters that generated it. This is the case for the CMB primary anisotropies, which have a nearly Gaussian distribution. However, non-linear gravitational evolution induces deviations from Gaussianity at small scales, leading to structures such as halos and voids. A large fraction of this non-Gaussian information cannot be captured by the power spectrum and requires including higher-order statistics (HOS), such as the bispectrum \citep{Scoccimarro:2000sn, Sefusatti:2006pa}. Despite being powerful in constraining cosmology \cite[e.g.][]{Chudaykin:2019ock, Changhoon_2019, Changhoon_2020, Chen:2021vba, Samushia_2021, Gualdi_2021},
extracting the information from HOS, however, is more challenging than from the power spectrum, due to complexities in their theoretical modeling, as well as significant computational cost (due to high dimensionality of the observables) of evaluating the signal and determining the expected noise \cite[see][for some of the recent applications of HOS to galaxy data]{Gil-Marin:2016wya,Philcox:2021kcw,Cabass:2022wjy,DAmico:2022gki}. This has spurred numerous works to construct optimal and computationally efficient estimators or data compression methods to extract the non-Gaussian information from the LSS  \citep[e.g.][]{Regan:2011zq, Obreschkow:2012yb,Schmittfull:2014tca,Chiang:2015pwa,Alsing:2017var, Gualdi:2018pyw, MoradinezhadDizgah:2019xun, Banerjee_2019, Gualdi:2019ybt, Heavens:2020spq, Schmittfull:2020hoi, Dai_2020, Massara_2020, Philcox:2020zyp, Uhlemann_2020,Naidoo2020, Banerjee_2020,Banerjee_2021,Gualdi:2022kwz}. 
In the context of constraining neutrino masses, the information contents of several of these methods, as well as standard bispectrum, were investigated on simulated data, but their potential has yet to be quantified on  observational data. 
Another powerful statistic is the marked power spectrum  \citep[see e.g.][]{Massara_2020}, which consists of the power spectrum of a so-called \textit{marked} field, computed from the original matter field to highlight aspects such as voids. Since it provides strong cosmological constraints and involves computing modified fields using exponents, we will use this statistic as one of the baselines for comparison.

Recently, promising results have been achieved using statistics derived from wavelet-based methods, such as the scattering transform and wavelet phase harmonics, which rely on cascades of wavelet transforms alternated with pointwise non-linearities, such as the complex magnitude (for scattering transforms) or ReLU and explicit phase modifications (for wavelet phase harmonics). 
Scattering transforms are capable of extracting higher-order statistical moments from signal while exhibiting robustness to small deformations (see Section \ref{sec:wavelets}).
The theoretical principle developed in \cite{mallat_group_2012} was initially applied to natural images \citep{bruna_invariant_2013,oyallon_deep_2015,oyallon_scaling_2017}, textures \citep{sifre_rotation_2013}, audio signals \citep{anden_deep_2014,anden_joint_2019,lostanlen_deep_2017}, and biomedical data \citep{chudacek_scattering_2014, chudacek_low_2014, villoutreix_synthesizing_2017, warrick_arrhythmia_2020} but was soon also shown to yield state-of-the-art results on 3D molecular electronic densities \citep{hirn_wavelet_2017, eickenberg_solid_2018, eickenberg_solid_2018-1} and a variety of astrophysical applications \citep{allys_rwst_2019,regaldo-saint_blancard_statistical_2020,cheng__new_2020,cheng_weak_2021,saydjari_classification_2021,valogiannis2021optimal}.
Wavelet phase harmonics were introduced in \citep{mallat_phase_2020} and have been shown to capture important statistical relations in fluid dynamic turbulence images \citep{zhang_maximum_2021} and explain phenomena observed in deep learning \citep{zarka2021separation}. 
They have also been successfully applied to perform statistical modelling as well as cosmological parameter estimation from 2D LSS fields~\citep{allys_new_2020,villaescusa-navarro_quijote_2020}, and statistical denoising of polarized dust observations~\citep{regaldo-saint_blancard_new_2021}.
While convolutional neutral networks (CNNs) have been shown in simulations to be a successful approach to extracting non-Gaussian information and constrain parameters from large-scale fields~\citep[e.g.][]{2021arXiv210910360V,2021arXiv210909747V,2020MNRAS.494.5761H,2019MNRAS.483.2524H,2020MNRAS.494..600M,2018PhRvD..97j3515G,2019NatAs...3...93R,2019A&C....2800307C,2021ApJ...916...42W}, these wavelet-based methods have been shown to achieve the performance of the state-of-the-art CNNs~\citep[e.g.][]{cheng__new_2020} in constraining parameters in the 2D setting. In addition, CNNs are usually described as black-box models which are very challenging to interpret. Wavelet-based approaches offer a window for interpretation.

A related topic predating scattering moments is that of wavelet moments. These consist in spatially integrating filtered signals after computing their absolute value and raising it to an exponent.
This
computes 
the absolute
moments of the marginal 
distribution of values
of the filtered signal.
These descriptors have been used successfully to analyze statistical properties of random fields and processes,
especially those with multifractal structure. 
Among many other applications, wavelet moments have been used to analyze the regularity of fluid dynamic turbulent flows \citep{Farge1992} and to determine the regularity, in terms of H\"older exponent, of multifractal signals \citep{MuzyBacryArneodo1991}. Many signals exhibit such multifractality. The prototypical multifractal signal is fractional Brownian motion \citep{jaffard2019}. It has also been shown that human brain M/EEG data can exhibit multifractal patterns with parameters depending on brain state \cite{larocca2021}.
For Gaussian fields and processes, the
wavelet moments become very simple functions of the exponent (see section \ref{sec:analytic_properties}).

In this paper we use 3D wavelet moments
as summary statistics for 3D cosmological fields obtained from the Quijote simulations \cite{villaescusa-navarro_quijote_2020}
and quantify the amount of
information they extract 
regarding cosmological parameters. 
The use of wavelets allows the creation of a multi-scale signal representation that (unlike e.g. the Fourier transform) preserves locality and can be made to vary smoothly with local changes in the signal. Since our summary statistics are based on these wavelet transforms, they inherit these properties.
Using appropriate wavelets and aggregation procedures,
we ensure that these descriptors are invariant to rotation in addition to the natural translation invariance that the summation provides.

We show that a very important aspect to achieving good constraints is the set of exponents $q$ to which the filtered fields are raised before integration. 
Indeed, exponent 2 will only contain information equivalent to power spectrum (see section \ref{sec:analytic_properties}), whereas other exponents, such as 1 (which in relation to exponent 2 can capture sparsity level) and other values will contribute significant additional information
about the cosmological parameters.
Furthermore, we introduce band-limited polar-separable wavelets of different types -- isotropic, oriented, and harmonic.
Band limitation is important for the analysis of simulation data, because cosmological N-body simulations are only accurate up to a certain resolution. We evaluate the constraining power of each of them, and combinations thereof. 
The resulting descriptors achieve state-of-the-art constraints on several cosmological parameters, improving on the power spectrum by factors 5--10. 

This paper is organized as follows. In section \ref{sec:wavelets} we present the wavelets and wavelet moments.
In section \ref{sec:methods} we described the data used and the method employed to evaluate the information content. In section \ref{sec:results} we present the main results. We draw the main conclusions of this work in section \ref{sec:conclusions}.

\section{Wavelets and wavelet moments}\label{sec:wavelets}

While the (isotropic) power spectrum (see section \ref{sec:pk}) is an excellent tool which accurately captures global frequency content of the density field at different scales, it has certain important drawbacks. First, this statistic is not stable to small deformations. Indeed, when the field is slightly deformed (e.g. $\delta(x)$ becomes $\delta(\tau(x))$ with $\tau$ close to the identity and its ``size'' $|\nabla\tau| < 1$ small), the high-frequency part of the power spectrum can vary greatly, out of proportion with the size of the deformation. The higher the frequency, the more pronounced this effect can be \citep{mallat_group_2012}. This property can be a contributing factor to larger observed error bars in power spectrum estimates for higher frequencies.
Second, the power spectrum is not able to characterize the presence of organized structures, like filaments, whose presence is encoded in particular in the phase of the Fourier transform.
Recovering this type of information requires capturing it in some form, for example with three-point correlation functions \citep{Obreschkow:2012yb}.
Contrary to this, wavelet transforms can be used to analyze a signal locally in both the spatial and frequency domains, allowing it to characterize local structures in a field. They exhibit equivariance to small deformations due to their localization and scaling properties and can therefore be made stable to such deformations \citep{mallat_group_2012}. 

A wavelet is an oscillatory function with zero mean that is localized around the origin. Wavelet signal analysis consists in convolving the signal with a family of wavelets. Such a family of wavelets can be derived from a single \textit{mother wavelet} $\psi$, by dilating with different dilation factors $2^j$:
\begin{equation}
        \psi_j(x) = 2^{-jd}\psi(2^{-j}x),
\end{equation}
where $d$ is the dimension of the signal and $j$ is the scale parameter. Scaling can be continuous, but here we use a finite set of exponentially increasing scales. \textit{Octave scaling} means that the wavelet size doubles with each scale index. For our applications we also consider sub-octave scaling controlled by a \textit{quality factor} $Q$.\footnote{The quality factor is originally defined as the ratio between peak frequency and bandwidth of a filter. A higher quality factor requires more filters because their bands are narrower. Here we simply use $Q$ to indicate the number of filters per octave.} The wavelets then become $\psi_j(x) = 2^{-jd/Q}\psi(2^{-j/Q}x)$. 
The total number of scales $J$ depends on the size of the signal and aliasing constraints: How much can the wavelet be dilated such that it still fits inside the box? The mother wavelet is the smallest wavelet and usually analyzes at the scale of the discretization step. An upper bound on the possible number of octave scales is then $\log_2 N$, where $N$ is the box size. Since we work with box sizes of $256^3$, but $k_\textrm{max}=0.5\ h/{\rm Mpc}$ corresponds to a radius of around 80 voxels in Fourier space, we choose the maximum number of octaves to be $J=6$ for all our analyses.
\footnote{
We will observe later that this number of scales is generous and the largest two scales are superfluous, making a reduction to $J=4$ possible. Because this is not a priori clear, we show it using feature reduction in section \ref{sec:feature_reduction}.}
   
To analyze a signal of dimension greater than one, we may additionally define an angular component of the wavelets to extract information pertaining to specific oriented frequencies. To analyze 3D matter density fields, we use three families of wavelets: isotropic, oriented, and harmonic. The oriented wavelets can be created from an oriented mother wavelet, by rotating it to different orientations, in addition to dilating it. The harmonic wavelets, on the other hand, are unoriented, but are able to sample different angular frequencies using spherical harmonics. For oriented and harmonic wavelets, we use a second index, $\lambda$, which denotes the orientation vector or polar angles of the wavelet direction in the case of oriented ones, and the spherical harmonic oscillation mode $(\ell, m)$ for the harmonic ones.
The isotropic wavelets are fully defined by their 1D radial component only.

For a signal $\delta(x)$, the wavelet transform is the set of filtered signals
\begin{equation}\label{eq:wavelet-transform}
W\delta = \{\delta\ast\psi_{j,\lambda}\}_{j\in J, \lambda\in\Lambda},
\end{equation}
where $\ast$ denotes convolution and $\Lambda$ denotes a collection of parameters related to angular frequency or orientation. In this work most of the wavelets are complex, and therefore have both a real and imaginary components. The same is thus true of the wavelet transform $W\delta$.

\subsection{Wavelet moments}

Wavelet transforms are \textit{equivariant} %
with signal translation,
and can also be made equivariant to other transformation groups.
In other words, a group action (such as a translation) applied to a signal results in a related group action being applied to its wavelet transform.
For oriented wavelets, a careful choice of orientations results in
exact rotation equivariance for chosen subgroups of $\SO(3)$,
or approximate equivariance on all of $\SO(3)$. 
A translation-invariant statistic can be created from a translation-equivariant one by aggregating it over space.  
A stable spatial aggregation mechanism is the spatial average. 
Similarly, using oriented wavelets, a rotation-invariant statistic can be obtained by aggregating over orientations. Again, this could be a simple average over the orientation parameters.

Since the integral of a wavelet-transformed signal is zero (because the wavelet has mean zero) and hence uninformative, we propose instead to spatially integrate powers of its complex modulus. The complex modulus of a wavelet transform corresponds to the first-order wavelet propagator used in scattering transforms \citep{mallat_group_2012}, which is why we denote it with the operator~$U$:
      \begin{equation}
           U[j, \lambda]\delta(x) = |\delta\ast\psi_{j, \lambda}(x)|.
           \label{eq:U}
      \end{equation}
For $q > 0$, the simplest types of wavelet moments are defined as spatial integrals of the wavelet modulus raised to the power $q$:
    \begin{equation}
        S_1[j, \lambda, q]\delta = \int_{\mathbb R^d}\left[U[j, \lambda]\delta(x)\right]^q\textrm dx.
        \label{eq:S1}
    \end{equation}
We adopt the notation $S_1$ for these descriptors from previous papers applying scattering transforms to cosmology (see e.g. \cite{allys_rwst_2019}), because they correspond to first-order scattering coefficients for $q=1$.
These descriptors have the desired translation invariance, but also encode non-trivial information about the structure of the field $\delta$.

Some of the wavelets introduced in this paper will require adjustments or further steps compared to (\ref{eq:U}) and (\ref{eq:S1}). These are detailed together with the wavelet definition below.

\subsection{Polar-separable spectral bump wavelets}

Wavelets can be defined in the spatial domain or the Fourier domain. 
Our wavelets are defined in the Fourier domain, because it allows straightforward handling of hard band-limit requirements that  are required for this work. Indeed, the Quijote simulations accurately capture effects below a cutoff frequency of 0.5~h/Mpc.
\footnote{ with a box size of 1~Gpc/$h$ and the typical resolution of $256^3$ voxels, the isotropic Nyquist frequency allows for resolution up to $128\frac{2\pi}{1000\textrm{Mpc}/h}\approx 0.8~h/\textrm{Mpc}$, which our band-limited wavelets can avoid using.}

In order to build rotational equivariance, 
we also must parameterize the angular component of the wavelet.
This suggests a spectral and polar-separable approach to wavelet construction. We start with the factorization 
\begin{equation}
    \hat\psi({\bf k}) = R(k)\,\alpha(\vartheta({\bf k}), \varphi({\bf k})),
\end{equation}
where ${\bf k}=(k_x, k_y, k_z)$ is the wave vector, and the magnitude $k = \sqrt{k_x^2 + k_y^2 + k_z^2}$, polar angle $\vartheta({\bf k}) = \arccos(k_z/k)$, and azimuthal angle $\varphi({\bf k}) = \arctan_2(k_y, k_x)$ are the spherical coordinates of $\bf k$. 
Observe that this factorization decouples scaling from rotation: a scaling acts only on the argument of $R$ and a rotation $T\in \SO(3)$ 
acts only on the argument of $\alpha$.

\begin{figure*}
    \centering
    \includegraphics[width=.99\linewidth]{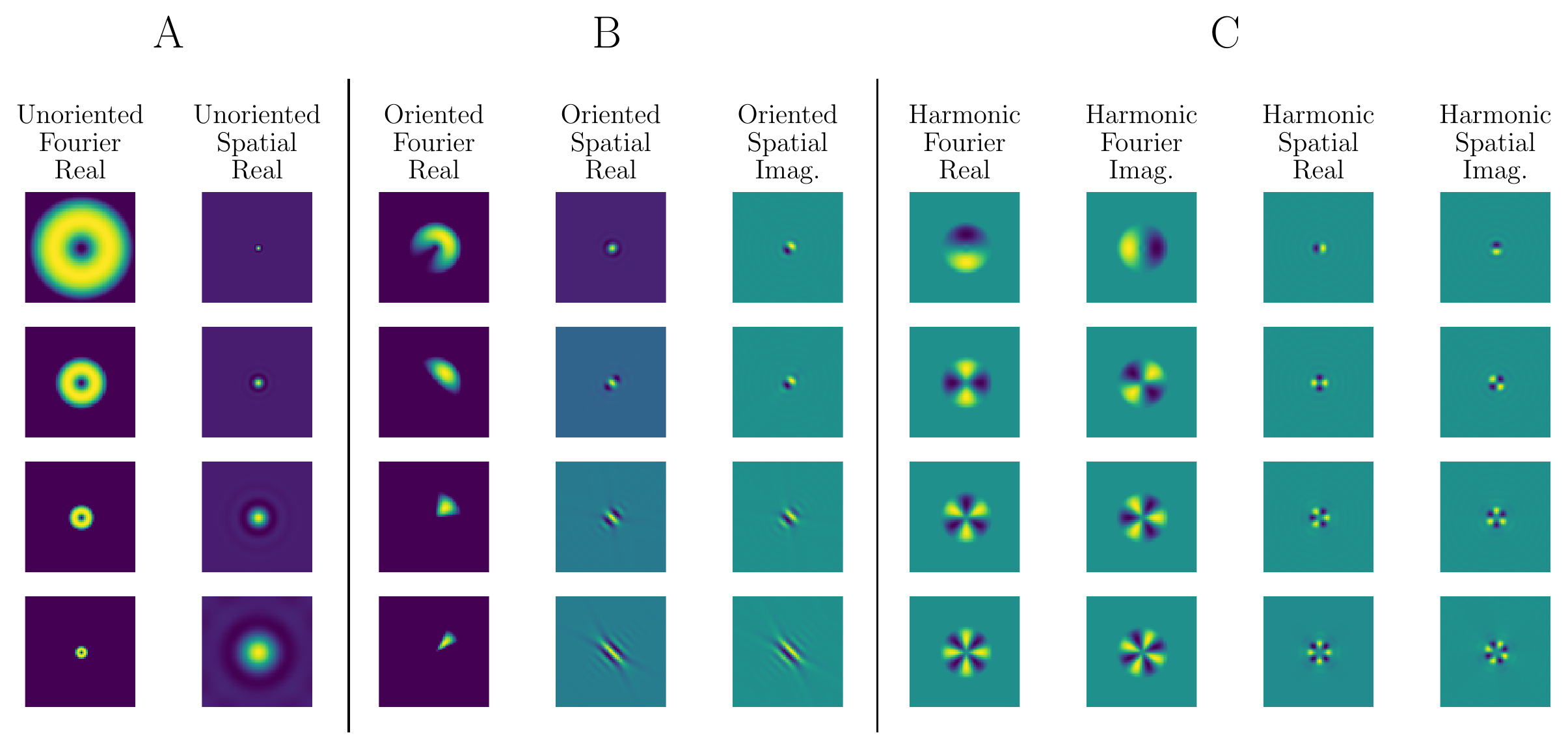}
    \includegraphics[width=.99\linewidth]{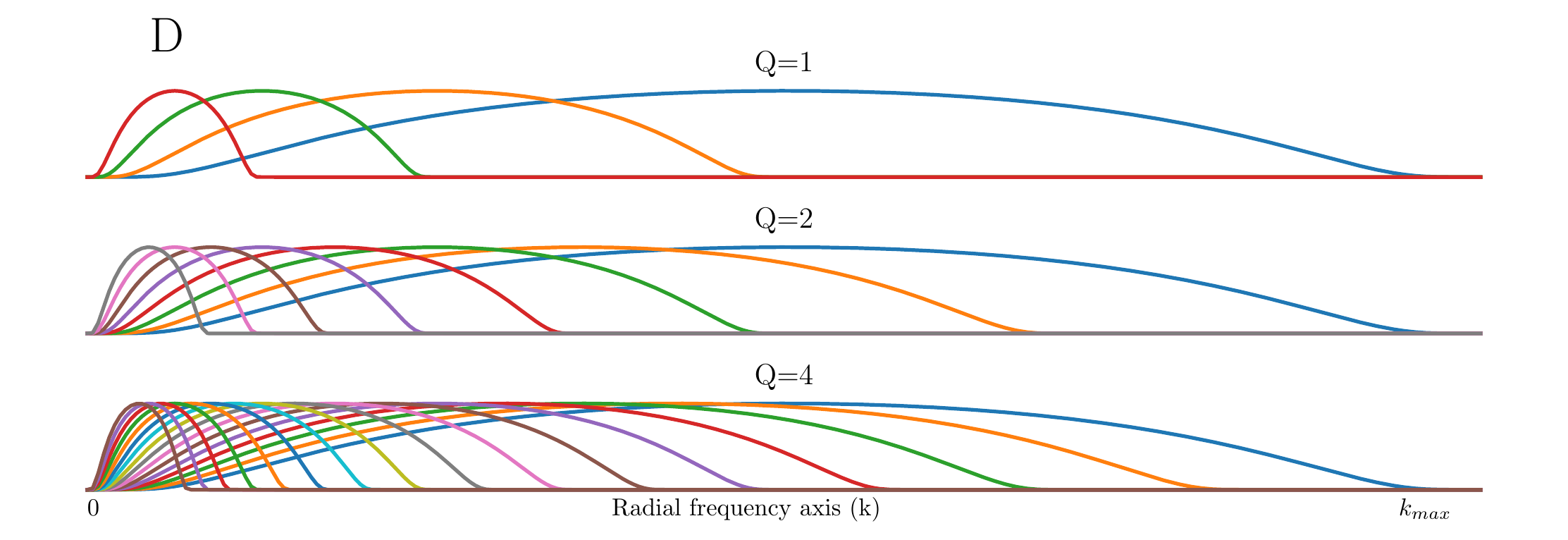}
    \caption{Visualization of slices through different parameterizations of the bump wavelet. The top panel shows 2D visualizations where the $z$-value is held constant. The columns show Fourier transform, spatial-domain real part, and imaginary part, respectively, of the same wavelet.
    Where only real parts are shown, the imaginary part is zero.
    The left-hand two columns (A) show several scales (small to large) of the isotropic bump wavelet, where the size of the wavelet doubles at each scale. This scaling progression is called octave scaling and corresponds to $Q=1$.
    The middle three columns (B) show bump wavelets with an orientation modulation, which induces orientation selectivity. Different orientation selectivities from $\pi$ to $\pi/8$ are shown.
    The last four columns (C) show bump wavelets with harmonic angular modulation. Different angular frequencies ($\ell=1$ to $\ell=4$) are shown.
    The three line plots in (D) show 1D radial slices through the central axis of an active part of each wavelet, which corresponds to a scaled version of the function $R(|k|)$. The top plot shows octave scaling ($Q=1$), the middle plot shows half-octave scaling ($Q=2$) and the bottom plot shows quarter-octave scaling ($Q=4$). }\label{fig:wavelets}
\end{figure*}

To define the radial part $R$, we create a smooth bump function with compact support. Smoothness ensures rapid decay of the wavelet envelope in the spatial domain. Compactness ensures radial band-limitedness of the wavelet in the frequency domain. The function
    \begin{equation}
        \beta(y) = \left\{\begin{array}{ll}\exp\left(-\frac{y^2}{1 -y^2}\right), & \textrm{if }|y| < 1, \\ 0, &\textrm{otherwise, }\end{array}\right.
    \end{equation}
satisfies these criteria. This function is also used in \cite{mallat_phase_2020}, \cite{zhang_maximum_2021}, and \cite{allys_new_2020}.
Given a maximum allowed frequency $k_\textrm{max} > 0$, we thus define
    \begin{equation}
        R(k) = \beta\left(\frac{k - k_\textrm{max} / 2}{k_\textrm{max} / 2}\right),
    \end{equation}
which has its peak at $k_\textrm{max} / 2$ and decays to zero at the origin and at $k_\textrm{max}$ (see bottom part of Fig. \ref{fig:wavelets} for its shape).

For the angular part $\alpha$, as noted earlier, we investigate three different choices $\alpha^\text{I}, \alpha^\text{O}$, and $\alpha^\text{H}$, corresponding to \textit{isotropic}, \textit{oriented}, and \textit{harmonic} wavelets. They allow to construct three different wavelet moments defined below in Eqs.~\eqref{EqDefWMIso},~\eqref{EqDefWMOriented}, and~\eqref{EqDefWMHarmonic}, respectively.

\paragraph{Isotropic wavelets}
For the isotropic wavelet, we simply set $\alpha^\text{I}(\vartheta, \varphi) = 1$ everywhere. Hence we obtain
\begin{equation}
    \hat\psi({\bf k}) = R(k).
\end{equation}
This results in a radially symmetric shell-like wavelet (Fig. \ref{fig:wavelets} top, left two columns). 
In the spatial domain it looks similar to a 3D Mexican hat wavelet (Laplacian of Gaussian). It performs local smoothing with a bell shape up to a scale-related characteristic radius, followed by a suppressive ring at a slightly larger radius, thus capturing local information in an isotropic manner.
With $U[j]\delta = |\delta\ast\psi_j|$ (mirroring (\ref{eq:U}), but without a supplementary orientation parameter), the corresponding wavelet moment (analogous to (\ref{eq:S1})) becomes
\begin{equation}
\label{EqDefWMIso}
    S_1^\text{I}[j, q]\delta = \int_{\mathbb R^3}[U[j]\delta(x)]^q\textrm dx.
\end{equation}

In our analyses we will use isotropic wavelets of different quality factors $Q=1,2,3,4$.

\paragraph{Oriented wavelets}
The oriented wavelet restricts the support of the angular part (and therefore of the wavelet) to make it directionally selective. We define the $z$-oriented wavelet with wave orientation along the Cartesian unit vector $\bm e_z$ and obtain the other directions by rotation. For an angular width $a > 0$, we write
    \begin{equation}
        \alpha^\text{O}_{\bm e_z}(\vartheta, \varphi) = \beta\left(\frac{\vartheta}{a}\right).
    \end{equation}
We choose the same bump function $\beta$ to specify the angular support of our oriented wavelet.
The wavelet is then defined as
\begin{equation}
\hat\psi({\bf k}) = R(k)\alpha^\text{O}_{e_z}(\vartheta, \varphi).
\end{equation}
For a collection of orientation vectors $\Lambda = \{e_\lambda\}_\lambda$ we can define a full set of oriented wavelets by rotating our z-oriented wavelet to each $e_\lambda$ direction. Indeed, for a rotation matrix $T_\lambda\in\SO(3)$ that verifies $T_\lambda e_z = e_\lambda$, we obtain
\begin{equation}
    \hat\psi_\lambda({\bf k}) = \hat\psi(T_\lambda^{-1}{\bf k}) = R(k)\alpha_{e_z}^O(\vartheta(T_\lambda^{-1}{\bf k}), \varphi(T_\lambda^{-1}{\bf k})).
\end{equation}

This definition works for any rotation $T_\lambda$ satisfying $T_\lambda e_z = e_\lambda$ because $\hat\psi$ is axisymmetric, i.e. rotationally symmetric around its own axis (here the z-axis).
This means that any rotated version of $\psi$ (obtained by applying a rotation $T_\lambda\in \SO(3)$
) can be described by only the two angles of the new orientation vector $e_\lambda$ (instead of three if it was not invariant to rotations about its own axis). See Fig. \ref{fig:wavelets} panel B for 2D cuts through an oriented wavelet. Here the parameter set $\Lambda$ consists of a collection of such orientations $\lambda = (\vartheta, \varphi)$. 
In the present paper we choose wavelet orientations to lie on the octahedron, which has six vertices, organized in three antipodal pairs. We choose them to be the unit vectors on the cartesian axes.

The wavelet propagator and wavelet moment $U[j, \lambda]$ and $S_1[j, \lambda, q]$ are defined exactly as in (\ref{eq:U}) and (\ref{eq:S1}). However, while the wavelet moment $S_1$ is translation-invariant, it is not yet fully rotation-invariant, since a rotated field will make the wavelet of a specific orientation respond differently. Due to statistical isotropy of the field, differently oriented wavelets may still lead to similar wavelet moments. In order to create a fully rotation-invariant descriptor and reduce variance, we average over orientations to obtain

\begin{equation}
\label{EqDefWMOriented}
    S_1^\text{O}[j, q]\delta = \frac{1}{|\Lambda|}\sum_{\lambda\in\Lambda} S_1[j, \lambda, q]\delta.
\end{equation}

In our analyses, we shall use angular width $a=\pi/2$ (denoted as $n_a=1$ to indicate the use of a single angular width) or a collection of angular widths ranging on a log scale from $a=\pi/4$ to $a=\pi$ in steps of $\sqrt{2}$ (denoted as $n_a=5$, because this collection contains 5 items). See table \ref{tab:npnanh} for details.

\paragraph{Harmonic wavelets}
For the harmonic wavelets, we choose $\alpha^\text{H}(\vartheta, \varphi)$ to be a spherical harmonic function. The set of angular wavelet parameters $\Lambda$ then comprises all pairs $(\ell, m)$, where $-\ell\leq m\leq \ell$ and $\ell \leq L$, where $L$ is the chosen maximal rotational frequency. We thus have
    \begin{equation}
        \alpha^\text{H}_{\ell, m}(\vartheta, \varphi) = Y_{\ell, m}(\vartheta, \varphi).
    \end{equation}
For harmonic bump wavelets 
\begin{equation}
\hat\psi_{\ell, m}({\bf k}) = R(k)\alpha^\text{H} _{\ell, m}(\vartheta, \varphi),
\end{equation}
a rotation by $T\in \SO(3)$ results in a per-voxel unitary transformation (a so-called Wigner D-matrix).
Because of this property, we define the wavelet modulus slightly differently from above in order to be able to build a robust invariant descriptor.
We define the harmonic modulus
    \begin{equation}
        U_H[j, \ell]\delta(x) = \sqrt{\sum_{m=-\ell}^{\ell}|\delta\ast\psi_{j, \ell, m}(x)|^2},
    \end{equation}
which is covariant to rotations and translations in a trivial way, because the quantity in the sum is a Euclidean norm that is invariant under the aforementioned unitary transformation, leaving only the voxel shifts from the translations and rotations, but removing the complex behavior in the $(\ell, m)$ channel axis.
The corresponding wavelet moment,
\begin{equation}
\label{EqDefWMHarmonic}
    S_1^\text{H}[j, l, q]\delta = \int_{\mathbb R^3}U_H[j, l]\delta(x)\textrm dx,
\end{equation}
is responsible for removing the remaining translation equivariance which yields an invariant descriptor.

In our analyses using harmonic wavelets, we shall use wavelet moments with $\ell=1$ (denoted $n_h=1$) or a collection of wavelet moments $\ell=1,2,3,4$ (denoted $n_h=4$), see table \ref{tab:npnanh} for details.
Note that the isotropic wavelet corresponds to $\ell=0$ and is sometimes combined with the harmonic wavelets.

\subsection{Important properties of wavelet moments}\label{sec:analytic_properties}

The wavelet moments $S_1[j, \lambda, q]$ exhibit a lot of structure, some of which can be straightforwardly mathematically characterized. Details can be found in the literature for multifractal signal analysis with wavelets (see \cite{jaffard2019} for a recent review). Here we will focus on two properties: a) The link between wavelet moments of exponent $q=2$ and the power spectrum and b) the behavior of wavelet moments under the assumption that the filtered field is Gaussian.

\subsubsection{Wavelet moments for $q=2$ and power spectrum}\label{sec:q2pk}

By Parseval's identity, for any finite collection of wavelets $\psi_i$ and signal $\delta$, we have
\begin{align}
    \sum_i\int_{\mathbb R^3}|\delta\ast\psi_i|^2 =: \sum_i\|\delta\ast\psi_i\|^2 = 
    \sum_i\|\mathcal F(\delta\ast\psi_i)\|^2 = \\ \nonumber
    \sum_i\|\hat\delta\hat\psi_i\|^2 = 
    \sum_i\int_{\mathbb R^3}|\hat\delta\hat\psi_i|^2 = 
    \int_{\mathbb R^3}|\hat\delta|^2\sum_i|\hat\psi_i|^2~,
\end{align}
where $\|\cdot\|$ is the $L^2$ norm, defined in the first equality.
This means that if $\sum_i|\hat\psi_i|^2$ is isotropic (i.e. it only depends on $|{\bf k}|$), then it can be expressed as a linear combination of isotropic power spectrum coefficients. 
Given the radial shape of our wavelets, this will result in a wide-band power spectrum coefficient.
For every isotropic wavelet and full $\ell$-spaces of harmonic wavelets, the sum is indeed perfectly isotropic. For oriented wavelets the isotropy of the sum depends on the angular width and orientation sampling of the wavelet. If the wavelet is wide enough in angle for the orientation sampling to cover the spectrum approximately evenly, then the oriented wavelets, too, can lead to an isotropic square sum in Fourier space.

The fact that wavelet moments for $q=2$ can become a function of the isotropic power spectrum means that they cannot contain more information about the field. Our results will confirm this.

\subsubsection{Wavelet moments under the Gaussian assumption}
Wavelet moments analyze the marginal distribution of the filtered field, i.e. its 1D histrogram. Seeing the voxel values of the filtered field as unordered samples from a distribution, wavelet moments are nothing other than empirical absolute moments of exponent $q$ of this distribution.

When the filtered field $\delta \ast\psi_{j}$ under study is Gaussian (and stationary, i.e. its covariance kernel only depends on relative positions), then the associated wavelet moments are absolute moments of a Gaussian distribution.

Recall that the absolute moments $m(q, \sigma)$ of a centered 1D Gaussian have the form 
\begin{equation}
m(q, \sigma) = \mathbb E_{x\sim\mathcal N(0, \sigma^2)}[|x|^q] = 2^{q/2}\sigma^q\Gamma\left(\frac{q+1}{2}\right)/\sqrt{\pi}.
\end{equation}

For a real stationary Gaussian field $x\ast\psi_j$, an estimator for the marginal variance $\sigma^2$ is the wavelet moment of exponent $q=2$, $S_1[j, 2]$. For a sampled field in 3D of typical size (e.g. $256^3$ voxels) this estimator is virtually exact. 
This means that we can compute all the wavelet moments as a function of the one with exponent $q=2$.
As a matter of fact, after taking the logarithm and subtracting out $\log\Gamma\left(\frac{q+1}{2}\right)$ and $-\log\sqrt{\pi}$, the moments become a linear function of $q$.
A field which does not exhibit this linear relation cannot be Gaussian.
Finally, note that by \ref{sec:q2pk}, the Gaussian wavelet moments then solely depend on the power spectrum. This is a restatement of the fact that all moments of a Gaussian are fully characterized by its covariance.

\begin{figure*}
    \includegraphics[width=.99\linewidth]{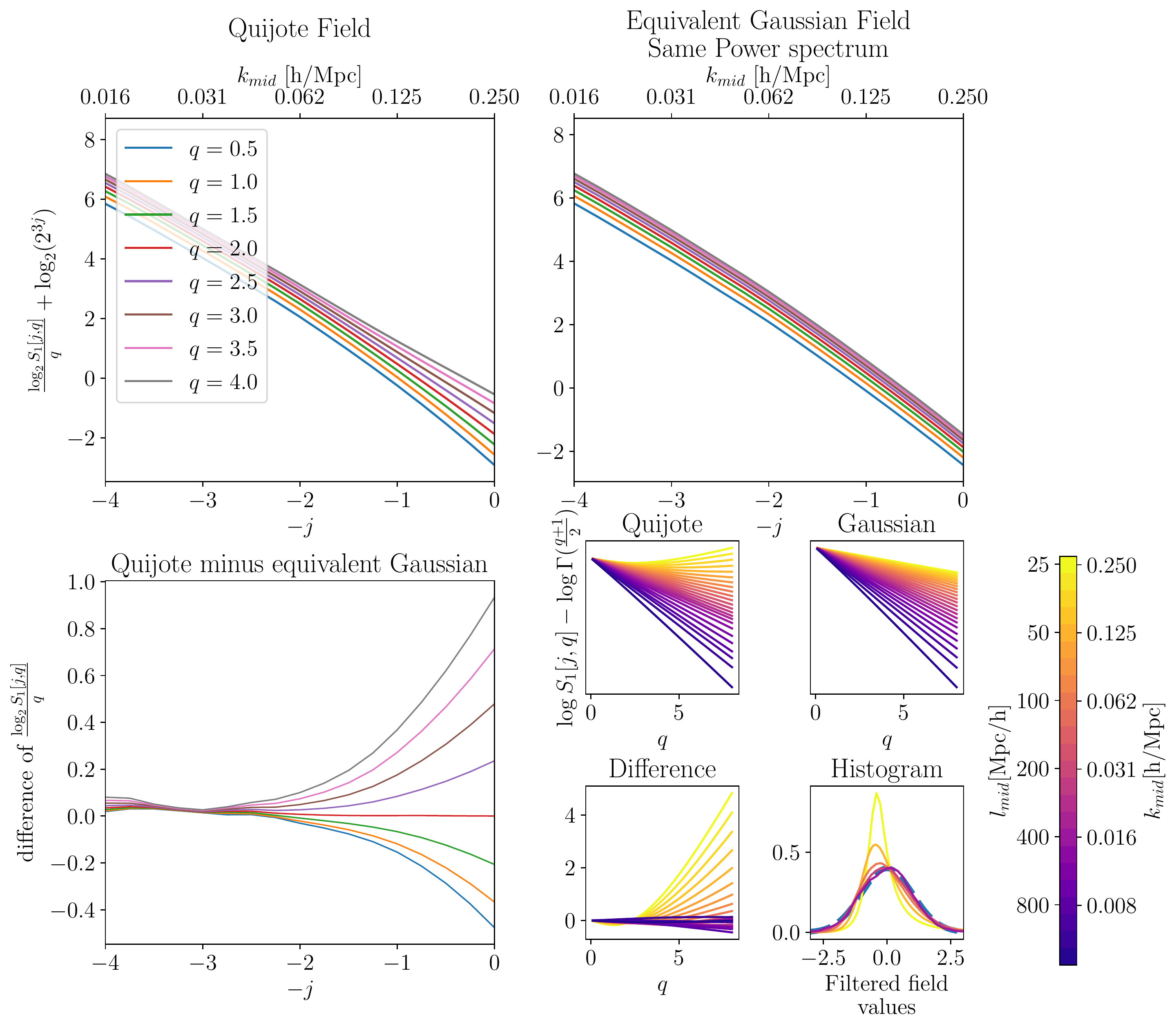}
    \caption{Visualization of the invariant wavelet moments for isotropic wavelets.
    {\it Top two panels:} Evolution of different wavelet moment exponents across scales for Quijote (left) and Gaussian field with same power spectrum (right). Note that Quijote moments fan out towards small scales ($-j > -2$) whereas the Gaussian ones do not. Moment $q=2$ is identical because it is determined by the power spectrum.
    The quasi-linear relation of wavelet coefficients as a function of scale is a hallmark of scale invariance.
    {\it Bottom left:} Difference of Quijote and Gaussian moments. Note that exponent $q=2$ traces a horizontal line at 0, smaller exponents curve down, larger exponents curve up.
    {\it Small panels:} Wavelet moments
    as a function of integral exponent, normalized by $\Gamma\left(\frac{q+1}{2}\right)$. A Gaussian field (right) shows a linear relation in log space, whereas the Quijote field (left) shows a linear relation only at large scales (dark colors), and a nonlinear relation at smaller scales (light colors). The nonlinear relation shows the scales at which the filtered field is not Gaussian.
    Bottom left panel shows the difference between these two.
    Bottom right panel shows the histogram of standardized filtered field at different scales. Standard Gaussian density in dashed lines overlays with large scales.}\label{fig:quijote_vs_gaussian}
\end{figure*}

See Figure \ref{fig:quijote_vs_gaussian} (small boxes) for a comparison between wavelet moments of a fully Gaussian field and wavelet moments of the Quijote data set. While for the fully Gaussian field the normalized wavelet moments are linear in log space, the Quijote field shows linearity only for the large scales, while smaller scales exhibit nonlinearity. This confirms that the small scales of these fields are not Gaussian. Further, the bottom right of the small panels shows histograms of the (standardized) filtered fields at different scales, which strongly skew away from Gaussian at small scales, and approach a Gaussian distribution at larger scales.

\subsection{Self-similarity, scale invariance and the decay of wavelet moments}
Self-similar signals are functions such that their behavior on a certain segment may be described by a dilated and shifted version of some other segment of the function. For example, a zoomed-in version of the function may be equal to a zoomed-out version at a different location. 
Scale invariance is a related concept where zoomed versions of a signal may leave certain quantities invariant, such as the overall statistics of a field. A scale-invariant function for all zooms $s$ obeys a relation of the form
\begin{equation}
    f(sx)\sim s^Df(x)
\end{equation}
for some $D > 0$. This relation leads to a power law in its Fourier transform. Such a power law can be picked up by wavelets. When using logarithmic scaling of the wavelets, the logarithm of the wavelet moments becomes a linear function of scale. This effect is at least approximately observable in Figure \ref{fig:quijote_vs_gaussian} (top two panels). It is furthermore known that between certain scales the Quijote fields indeed follow a power law.

\section{Fisher information on cosmological parameters}\label{sec:methods}

\subsection{Simulated data and analysis methodology}
In this work we quantify the information content of a given statistic about the cosmological parameters using the Fisher matrix formalism \citep{Tegmark_1997} on the Quijote simulations \cite{villaescusa-navarro_quijote_2020}. 

\subsubsection{Fisher formalism}
 The Cram\'er--Rao bound 
 is a lower bound on the error of an unbiased estimator $\hat\theta(y)$ of a parameter $\theta$ of a parametric distribution $p(y|\theta)$.
 It states that
 \begin{equation}
     \textrm{cov}(\hat\theta)\succeq F^{-1}(\theta)
 \end{equation}
 where $F(\theta)$ is the Fisher information matrix and the inequality is taken in terms of the partial ordering on positive-semidefinite matrices ($A\succeq B$ if for all $v$ we have $v^TAv \geq v^TBv$).
 The Fisher matrix is defined as the covariance of the score function: $F(\theta) = \textrm{cov}_p(\nabla_\theta p(\cdot| \theta))$ or equivalently its expected negative Hessian.
In particular, the Cram\'er--Rao bound implies that the marginal error in estimating the $i$th parameter $\theta_i$ is $\sigma(\theta_i)\geq\sqrt{(F^{-1})_{ii}}$.
When $y$ is a Gaussian-distributed statistic, whose mean $\mu$ depends on a set of parameters $\theta$ (and its covariance $\Sigma$ does not), the Fisher information takes the form 
 \begin{equation}
    F(\theta) = \nabla_\theta\mu^T\Sigma^{-1}\nabla_\theta\mu~,
    \label{eq:gauss-fisher}
\end{equation}
In this paper we make the standard assumption that it is sufficient to compute the Fisher matrix in this framework.
In this work we will use a large set of numerical simulations from the Quijote suite \citep{villaescusa-navarro_quijote_2020} to evaluate the two ingredients needed to compute the Fisher matrix in this form: the partial derivatives of the considered statistic with respect to the parameters, and the covariance matrix. 
We derive bounds for the standard power spectrum, the marked power spectrum, and the wavelet moments for the three families of the wavelets that we described earlier.

\subsubsection{Quijote data set}

The Quijote simulations are a suite of 44,100 full N-body simulations that span a wide range of values for 
seven
cosmological parameters, and was designed to quantify the information content on generic cosmological statistics, as well as to train machine learning models. At its core, Quijote provides simulations arranged to compute derivatives of generic statistics with respect to cosmological parameters and to evaluate their covariance matrix -- the two ingredients needed to calculate the Fisher matrix described above. 

The covariance matrix is computed using 15,000 simulations of a fiducial cosmological model with $\Omega_m = 0.3175$, $\Omega_b = 0.049$, $h = 0.6711$, $n_s = 0.9624$, $\sigma_8 = 0.834$ and $M_\nu = 0.0 \ eV$, where $\Omega_m$ and $\Omega_b$ are the energy density of matter and baryons respectively; $h$ is the reduced Hubble constant $h \equiv H_0 / 100$ with $H_0$ in units of ${\rm km \ s^{-1} \ Mpc^{-1}}$; $n_s$ is the spectral index, $\sigma_8$ is the present day linear theory root-mean-square amplitude of the matter fluctuation spectrum averaged in spheres of radius $8~h^{-1} \ {\rm Mpc}$, and $M_\nu$ is the sum of neutrino masses. 
The partial derivatives are evaluated using a set of 1,000 simulations where the value of a single cosmological parameter is varied at a time. We refer the reader to \citet{villaescusa-navarro_quijote_2020} for further details on the Quijote simulations.
Every simulation follows the evolution of $512^3$ dark matter particles in a periodic comoving volume of $(h^{-1}{\rm Gpc})^3$ from redshift $z=127$ down to $z=0$.
Here we use the simulations from a single redshift, $z=0$.

For each simulation we first compute the 3D matter field by depositing particle masses into a regular 3D cube with $256^3$ voxels employing the piecewise cubic spline (PCS) mass assignment scheme. Next, from the 3D grids we compute both the power spectrum and the wavelet modulus integral coefficients. Finally, we calculate the covariance matrix of the considered statistics and the partial derivatives and use these to compute the Fisher matrix, and the square root of the diagonal of its inverse, which leads us to the constraints we report. 
In order to ensure numerical stability of the inverse Fisher matrix estimate, we recondition the extracted statistics by standardizing and PCA-transforming. This procedure is described in detail in appendix \ref{sec:reconditioning}.

For the whole study, we consider two different fields: the total matter field (m), representing the sum of cold dark matter, baryons, and neutrinos, and the cold dark matter plus baryons field (cb). Note that when neutrinos have zero mass, both fields are the same. 
It is important to quantify the information content in both cases, as the m field can be surveyed using weak lensing while dark matter halos and galaxies are tracers of the underlying cb field \citep{Paco_mu1, Paco_mu2}.

\subsection{Alternative summary statistics}\label{sec:pk}

A ubiquitous statistic with considerable constraining power on cosmological parameters is the 
isotropic
power spectral density. For a density field $\delta(x)$ it is defined as 
        \begin{equation}
            P_k(\delta) =
            \frac{1}{4\pi k^2}\int_{|{\bf k'}| = k}|\hat\delta({\bf k'})|^2\textrm d{\bf k'}
        \end{equation}
which is the integral at constant spectral radius along the angular part of the Fourier transform of the two-point (or auto-correlation) function.
In practice, this quantity is computed on the discrete Fourier transform of a density field, aggregating power-spectrum energy in bins of width one, and averaging. Throughout this paper, we use this power spectrum as a baseline and as a complement for the other statistics we present.

The marked power spectrum \citep[see e.g.][]{Massara_2020} applies a nonlinear transformation to the density field before estimating its power spectrum. We first define the \textit{mark}: 
\begin{equation}
    m(\delta; \delta_s, r, p) = \left(\frac{1 + \delta_s}{1 + \delta_s + \delta_r(\delta)}\right)^p,
\end{equation}
where $\delta_s$ is a damping parameter, $p$ is an exponent, and $r >0$ is a radius indicating a smoothing scale. 
The quantity $\delta_r(\delta)$ is a smoothed 
version
of $\delta$, obtained by local smoothing with a filter $\phi_r$ (usually a top-hat filter) of radius $r$, as 
$\delta_r(\delta) = \delta\ast\phi_r$. 
The marked power spectrum is then defined as
\begin{equation}
    M_k(\delta; \delta_s, r, p) = P_k(m(\delta; \delta_s, r, p) \delta),
\end{equation}
i.e. it consists in the computation of isotropic power spectrum coefficients on the marked density field,
which is the product of the mark and the density field.
Since the marked density field highlights different properties from the original density field, such as voids, the properties captured by the marked power spectrum are different from those captured by the power spectrum applied to the original density maps.

We will use the power spectrum and the marked power spectrum as baselines for comparison.
The marked power spectrum is a useful additional baseline because of its strong performance and the field modifications using exponents, which we also study here.
For the marked power spectrum we will use the parameter settings specified in \citep{Massara_2020} ($\delta_s = 0.25, r = 10h/Mpc, p = 2$ for $M$) and report the numbers from this publication in our results table.

\section{Results}
\label{sec:results}

\subsection{Constraining power of wavelet modulus integrals}
To assess the information content of wavelet moments for constraining cosmological parameters, we perform series of Fisher forecasts, progressively increasing the number of included statistics.

The results can be summarized as follows.
\begin{itemize}
    \item Wavelet moments of exponent 2 using isotropic, harmonic, and most oriented wavelets, leads to constraints similar to those obtained by the isotropic power spectrum $P_k$. This is due to Parseval's identity (see section \ref{sec:analytic_properties}).
    \item Isotropic wavelets alone, using integral exponents of 1 and lower in addition to exponent 2 lead to very strong performance improvements over the isotropic power spectrum (factor $\sim 5$).
    \item Oriented wavelets 
    using
    integral exponents 1 and lower in addition to exponent 2 yield an improved constraint over isotropic wavelets, especially for neutrino masses.
    \item Using harmonic wavelets leads to further improvements in the neutrino mass constraint.
\end{itemize}
We address each of these items in this section. Exhibiting them will involve presenting results tables which show wavelet moment configurations and constraints. The number of integral powers is abbreviated to $n_q$, the number of angular widths for oriented wavelets to $n_a$, and the number of harmonic frequencies to $n_h$. These numbers correspond to specific choices, which are mapped in table \ref{tab:npnanh}.

\begin{table}[]
\centering
\begin{tabular}{|c|c|c|c|}
\hline
                    & $n_q$                                         & $n_a$                           & $n_h$      \\ \hline
1                   & 2                                             & $\pi/2$                         & 1          \\ \hline
2                   & 1, 2                                          &                                 &            \\ \hline
4                   &                                               &                                 & 1, 2, 3, 4 \\ \hline
\multirow{2}{*}{5}  & \multirow{2}{*}{1/4, 1/2, 1, 2, 4}            & $\pi$, $\pi/\sqrt{2}$, $\pi/2,$ &            \\
                    &                                               & $\sqrt{2}\pi/4$, $\pi/4$        &            \\ \hline
\multirow{2}{*}{11} & 1/8, $\sqrt{2}/8$, 1/4                        &                                 &            \\
                    & $\sqrt{2}/4$, 1/2, $\cdots$, 4 &                                 &            \\ \hline
\end{tabular}
    \caption{Mapping between values of $n_q, n_a, n_h$, used in all results tables, and the corresponding selections of values. The values of $n_\ast$ are indicated in the first column, and count the size of the collection of values used. The first row indicates that when $n_q=1$, only the exponent $q=2$ is used, when $n_a=1$ the angular width $a=\frac{\pi}{2}$ is used, and when $n_h=1$, then only the angular wave number $\ell=1$ is used. The second row indicates that when $n_q=2$, then the collection $\{1,2\}$ is used for the exponent $q$.}
    \label{tab:npnanh}
\end{table}

\subsubsection{$L^2$ wavelet modulus integrals and power spectrum}
In section \ref{sec:q2pk} we showed
that wavelet moments of power $q=2$ with isotropic spectral square sum cannot capture any information beyond the isotropic power spectrum.
We confirm this empirically 
in table \ref{tab:p2}, which can be summarized as follows:
\begin{enumerate}
    \item Using isotropic wavelets, the power spectrum constraints can be approached from above by increasing the quality factor $Q$ (rows 2--5). These wavelets essentially create wide-band power-spectrum statistics. Increasing $Q$ adds more wavelets and decreases the distances between the frequency centers of two adjacent wavelets. Though their bands are wide, especially in high frequency, this permits better and better recovery of the power spectrum.
    \item Combining isotropic wavelet modulus integrals of power 2 with isotropic power spectrum leads to no significant constraint improvement over power spectrum alone (row 6). This confirms that the wavelet modulus integrals of exponent 2 do not contain more information on the cosmological parameters than the power spectrum.
    \item All harmonic wavelet configurations (rows 8--9) as well as octahedron-sampled oriented wavelets of width $\pi$ (row 7) lead to the same constraints as isotropic wavelets of the same quality factor (row 5).
\end{enumerate}

\begin{table*}[]
    \centering
    \setlength\tabcolsep{2.2pt}
\begin{tabular}{c|c|c|c|c|c|c|c||c|c|c|c|c|c||c|c|c|c|c|c}
\multicolumn{8}{c}{} & \multicolumn{6}{c}{matter field} & \multicolumn{6}{c}{cb field}\\\hline
 {} &Desc& Q & $n_\textrm{p}$ & $n_\textrm{a}$ & $n_\textrm{h}$ &
 $n_\textrm{coef}$ & $n_\textrm{eff}$ 
 & $\Omega_m$ & $\Omega_b$ & $h$ & $n_s$ & $\sigma_8$ & $M_\nu$&
  $\Omega_m$ & $\Omega_b$ & $h$ & $n_s$ & $\sigma_8$ & $M_\nu$\\ 
 \hline\hline
1 & $P_k$ & {} & {} & {}& {}  & 79&79 & 0.098 &	0.039 &	0.51 & 	0.50 &	0.014 &	0.77& 0.070 & 0.018 & 0.19 & 0.14 & 0.12 &
       1.9\\\hline
       \hline
2 & $W_I$ & 1 & 1 & 0 & 0 & 6&6 & 0.35 & 0.24 & 2.8 & 2.1 & 0.052 & 2.7& 8.8 & 3.7 & 25 & 19 & 8.4 & 135.0\\\hline
3 & $W_I$ & 2 & 1 & 0 & 0 & 12&12 & 0.23 & 0.14 & 1.7 & 1.4 & 0.036 & 1.9& 0.14 & 0.039 & 0.44 & 0.25 & 0.22 & 3.6\\\hline
4 & $W_I$ & 3 & 1 & 0 & 0 & 18&18 & 0.18 & 0.086 & 1.1 & 1.0 & 0.027 & 1.5& 0.12 & 0.027 & 0.31 & 0.22 & 0.20 & 3.3\\\hline
5 & $W_I$ & 4 & 1 & 0 & 0 & 24&24 & 0.14 & 0.059 & 0.75 & 0.72 & 0.020 & 1.1& 0.10 & 0.024 & 0.27 & 0.20 & 0.18 & 2.9\\\hline
\hline
6 & $W_I+P_k$ & 4 & 1 & 0 & 0  & 103 & 101 & 0.094 & 0.037 & 0.49 & 0.48 & 0.013 & 0.74 & 0.076 & 0.018 & 0.20 & 0.15 & 0.13 & 2.2\\\hline
\hline
7 & $W_O$ & 4 & 1 & 1 & 0& 24 & 24 & 0.14 & 0.059 & 0.75 & 0.72 & 0.020 & 1.1& 0.10 & 0.024 & 0.27 & 0.20 & 0.18 & 2.9\\\hline
\hline
8 & $W_H$ & 4 & 1 & 0 & 1 &24 & 24 & 0.14 & 0.060 & 0.76 & 0.73 & 0.020 & 1.1 & 0.10 & 0.024 & 0.27 & 0.2 & 0.18 & 2.9\\\hline
9& $W_H + W_I$ & 4 & 1  & 0 & 4&120 & 24 & 0.14 & 0.060 & 0.76 & 0.73 & 0.020 & 1.1 & 0.10 & 0.024 & 0.27 & 0.2 & 0.18 & 2.9\\\hline

    \end{tabular}
    \caption{Constraints using wavelet modulus integrals with exponent 2. When $\sum_i |\hat\psi_i|^2$ is isotropic, these can never improve upon isotropic power spectrum. 
    Rows 2--5: Increasing Q allows approaching $P_k$ constraints from above.
    Row 6: Wavelets modulus integrals with $q=2$ do not add information beyond power spectrum.
    Rows 7--8: Oriented wavelets do not provide more information than isotropic wavelets if $\sum_i|\hat\psi_i|^2$ is isotropic (it is not for row 8).
    Rows 9--10: Harmonic wavelets do not contribute more information than isotropic wavelets at integral exponent 2.}
    \label{tab:p2}
\end{table*}

\subsubsection{Isotropic wavelets: Adding other exponents than $q=2$ greatly improves constraints}

By adding exponents other than 2 we leave the regime of Parseval's identity where sums of wavelet moments might collapse into simpler wavelet moments. 
In this new setting, the spatial summing procedure and the rotation-invariance-inducing operations do not commute anymore, and more non-redundant information can be extracted.

Results for this setting are in table \ref{tab:iso-powers}.
In an effort to remain concise, because $Q=4$ consistently leads to the best constraints, we report only results for $Q=4$ from now on.
The results can be summarized as follows.
\begin{enumerate}
    \item Adding integral power 1 in addition to the already used exponent 2 leads to a large improvement in constraining power, compared to $q=2$ only and power spectrum. We observe a $>3\times$ improvement over $P_k$ on every parameter using the $m$ field, and a $>30\%$ improvement over $P_k$ using the $cb$ field (row 4).
    \item Using more integral exponents, e.g. $1/4, 1/2, 1, 2, 4$ (row 5) yields another factor 2 improvement on $M_\nu$ constraint for $m$ field, and a slightly weaker improvement for $\sigma_8$ and $M_\nu$ using the $cb$ field.
    The integral exponents below 1 carry the bulk of the improvements. Exponents between 0 and 1 compress the non-zero part of the filtered field towards 1, making it closer to an indicator signaling presence or absence of filtered field, which one can interpret as a sparsity measure.
    Exponents above 1 (e.g. 4) highlight the peaks of the filtered fields, so the spatial average will tend towards counting peaks.
    \item Adding more integral powers ($1/8, \sqrt{2}/8,1/4,\dots,4$, row 6) does not significantly improve the constraints. 
    We observe that adding these exponents also recovers the power-spectral information lost to broad-band filtering in results table \ref{tab:p2}. This can be seen in row 7: Adding $P_k$ does not really improve the constraints.
    We arrive at a set of constraints comparable to marked power spectrum $M_k$ (row 7 vs.\ row 3).
\end{enumerate}

\begin{table*}[]
    \centering
    \setlength\tabcolsep{2.2pt}
\begin{tabular}{c|c|c|c|c|c|c|c||c|c|c|c|c|c||c|c|c|c|c|c}
\multicolumn{8}{c}{} & \multicolumn{6}{c}{matter field} & \multicolumn{6}{c}{cb field}\\\hline
 {} &Desc& Q & $n_\textrm{p}$ & $n_\textrm{a}$ & $n_\textrm{h}$ &
 $n_\textrm{coef}$ & $n_\textrm{eff}$ 
 & $\Omega_m$ & $\Omega_b$ & $h$ & $n_s$ & $\sigma_8$ & $M_\nu$&
  $\Omega_m$ & $\Omega_b$ & $h$ & $n_s$ & $\sigma_8$ & $M_\nu$\\ 
 \hline\hline
1 & $P_k$ & {} & {} & {}& {}  & 79&79 & 0.098 &	0.039 &	0.51 & 	0.50 &	0.014 &	0.77& 0.070 & 0.018 & 0.19 & 0.14 & 0.12 &
       1.9\\\hline
2 & $W_I$ & 4 & 1 & 0 & 0 & 24&24 & 0.14 & 0.059 & 0.75 & 0.72 & 0.020 & 1.1& 0.10 & 0.024 & 0.27 & 0.20 & 0.18 & 2.9\\\hline
3 & $M_k$ & {} & {} & {} & {} & 79&79 & 0.013 & 0.010 & 0.098 & 0.048 & 0.0019 & 0.017& 0.018 & 0.0099 & 0.092 & 0.045 & 0.030 & 0.50\\\hline
\hline
4 &$W_I$ & 4 & 2 & 0 & 0 & 48&48 & 0.014 & 0.011 & 0.10 & 0.058 & 0.0016 & 0.048& 0.044 & 0.013 & 0.13 & 0.081 & 0.084 & 1.3\\\hline
5 & $W_I$ & 4 & 5 & 0 & 0 & 116 & 120 & 0.014 & 0.011 & 0.097 & 0.055 & 0.0014 & 0.023& 0.026 & 0.011 & 0.10 & 0.063 & 0.044 & 0.71\\\hline
6 & $W_I$ & 4 & 11 & 0 & 0 & 264 & 127 & 0.013 & 0.010 & 0.093 & 0.054 & 0.0014 & 0.022& 0.023 & 0.010 & 0.097 & 0.058 & 0.039 & 0.623\\\hline
\hline
7 & $W_I+P_k$ & 4 & 11 & 0& 0  & 343 & 202 & 0.013 & 0.0088 & 0.085 & 0.049 & 0.0014 & 0.021& 0.023 & 0.0092 & 0.090 & 0.055 & 0.038 & 0.61\\\hline
\hline

    \end{tabular}
    \caption{Constraints using wavelet modulus integrals with multiple exponents. 
    Row 4: Adding exponent 1 yields vastly improved parameter constraints over $q=2$ (row 2) only and $P_k$ (row 1).
    Rows 5--6: Adding more integral exponents lead to further constraint improvements.
    Row 7: Power spectrum does not significantly improve row 6.}
    \label{tab:iso-powers}
\end{table*}

\subsubsection{Oriented and harmonic wavelets can further improve the neutrino mass constraint}
Oriented and harmonic wavelet modulus integrals give additional access to the filamentary structure of the matter field, by being able to characterize their locally anisotropic properties. This helps in particular for constraining neutrino mass, which has an influence on their evolution.
The results are presented in table \ref{tab:oriented-harmonic} and can be summarized as:
\begin{enumerate}
    \item Using oriented wavelets with one angular width ($\pi/2$) and 5 integral powers (row 4) leads to similar constraints for the $m$ field as the best isotropic setting (row 2) using half the number of coefficients. The neutrino mass constraint is slightly improved. Using 11 integral powers does not change the constraints.
    For the $cb$ field the constraints are worse than than those of the isotropic wavelets.
    \item Combining oriented wavelet modulus integrals with 5 different angular selectivities and isotropic wavelet modulus integrals (rows 6--8) as the non-angularly-selective limit achieves parameter constraints almost uniformly tighter or equal to isotropic-only (row 2) and the marked power spectrum (row 3) both for $m$ and $cb$ fields.
    \item Using harmonic wavelets of type $\ell=1$ (rows 9--10), similar bounds can be achieved to using one type of oriented wavelet (rows 4--5), with a significant improvement on neutrino mass constraints.
    \item Combining harmonic wavelets for $\ell=1,2,3,4$ and adding isotropic wavelets, which correspond to $\ell=0$, leads to greatly improved bounds on all parameters (rows 11--12). Combining these descriptors with power spectrum leads to the strongest set of constraints presented in this paper (row 13).
\end{enumerate}

\begin{table*}
\begin{center}
\setlength\tabcolsep{2.2pt}
\begin{tabular}{c|c|c|c|c|c|c|c||c|c|c|c|c|c||c|c|c|c|c|c}
\multicolumn{8}{c}{} & \multicolumn{6}{c}{matter field} & \multicolumn{6}{c}{cb field}\\\hline
 {} &Desc& Q & $n_\textrm{p}$ & $n_\textrm{a}$ & $n_\textrm{h}$ &
 $n_\textrm{coef}$ & $n_\textrm{eff}$ 
 & $\Omega_m$ & $\Omega_b$ & $h$ & $n_s$ & $\sigma_8$ & $M_\nu$&
  $\Omega_m$ & $\Omega_b$ & $h$ & $n_s$ & $\sigma_8$ & $M_\nu$\\ 
 \hline\hline
1 & $P_k$ & {} & {} & {}& {}  & 79&79 & 0.098 &	0.039 &	0.51 & 	0.50 &	0.014 &	0.77& 0.070 & 0.018 & 0.19 & 0.14 & 0.12 &
       1.87\\\hline
2 & $W_I+P_k$ & 4 & 11 & 0& 0  & 343 & 202 & 0.013 & 0.0088 & 0.085 & 0.049 & 0.0014 & 0.021& 0.023 & 0.0092 & 0.090 & 0.055 & 0.038 & 0.61\\\hline
3 & $M_k$ & {} & {} & {} & {} & 79&79 & 0.013 & 0.010 & 0.098 & 0.048 & 0.0019 & 0.017& 0.018 & 0.0099 & 0.092 & 0.045 & 0.030 & 0.50\\\hline
\hline

4 & $W_O$ & 4 & 5 & 1 & 0 & 120& 101 & 0.013 & 0.010 & 0.091 & 0.051 & 0.0014 & 0.019 & 0.028 & 0.011 & 0.10 & 0.060 & 0.050 & 0.79\\\hline
5 & $W_O$ & 4 & 11 & 1 & 0 & 264& 108 & 0.013 & 0.010 & 0.091 & 0.051 & 0.0014 & 0.019 & 0.027 & 0.011 & 0.098 & 0.058 & 0.047 & 0.76\\\hline
\hline

        6 & $W_O + W_I$ & 4 & 5 & 5&0  & 720 &414 & 0.012 & 0.0096 & 0.085 & 0.046 & 0.0014 & 0.016& 0.019 & 0.0098 & 0.088 & 0.050 & 0.029 & 0.46\\\hline
        7 & $W_O + W_I$ & 4 & 11 & 5&0& 1584 &439  & 0.012 & 0.0094 & 0.084 & 0.046 & 0.0014 & 0.016& 0.018 & 0.0096 & 0.086 & 0.048 & 0.026 & 0.42\\\hline
8 & $W_O + W_I + P_k$ & 4 & 11 & 5& 0 
& 1663 & 512 & 0.012 & 0.0082 & 0.077 & 0.043 & 0.0013 & 0.015& 0.018 & 0.0085 & 0.079 & 0.046 & 0.026 & 0.42\\\hline        
\hline

        9 & $W_H$ & 4 & 5 & 0 & 1 &120 & 98 & 0.012 & 0.0097 & 0.086 & 0.048 & 0.0014 & 0.016 & 0.025 & 0.010 & 0.092 & 0.053 & 0.044 & 0.70\\\hline
        10 & $W_H$ & 4 & 11 & 0 & 1 &264 & 104 & 0.012 & 0.0097 & 0.086 & 0.048 & 0.0014 & 0.016 & 0.024 & 0.010 & 0.092 & 0.053 & 0.043 & 0.69\\\hline
\hline

        11 & $W_H + W_I$ & 4 & 5  & 0 & 4&600 & 301 & 0.012 & 0.0090 & 0.079 & 0.044 & 0.0014 & 0.014 & 0.014 & 0.0089 & 0.078 & 0.043 & 0.016 & 0.26\\\hline
        12 & $W_H + W_I$ & 4 & 11  & 0 & 4&1320 & 308 & 0.012 & 0.0089 & 0.079 & 0.043 & 0.0014 & 0.014 & 0.014 & 0.0089 & 0.078 & 0.042 & 0.016 & 0.25\\\hline
    13 & $W_H + W_I + P_k$ & 4 & 11 & 0 & 4 & 1399 & 382 & 0.011 & 0.0078 & 0.072 & 0.040 & 0.0013 & 0.013 & 0.013 & 0.0078 & 0.072 & 0.040 & 0.016 & 0.25\\\hline

\end{tabular}

\caption{Constraints using wavelet modulus integrals with oriented and harmonic wavelets.
Rows 4--5: Oriented wavelets with angular width $\pi/2$ improve over isotropic wavelet modulus integrals.
Rows 6--8: Oriented wavelets with 5 different angular widths combined with isotropic wavelets (``infinite'' angular width) lead to constraint improvement over single angular width.
Rows 9--10: Harmonic wavelet moments with $\ell=1$ achieve similar constraints to oriented wavelets with fewer coefficients.
Rows 11--13: Harmonic wavelet modulus integrals with $\ell=0,1,2,3,4$ yield the best constraints.
}\label{tab:oriented-harmonic}
\end{center}
\end{table*}

\begin{figure*}
    \centering
    \includegraphics[width=.49\linewidth]{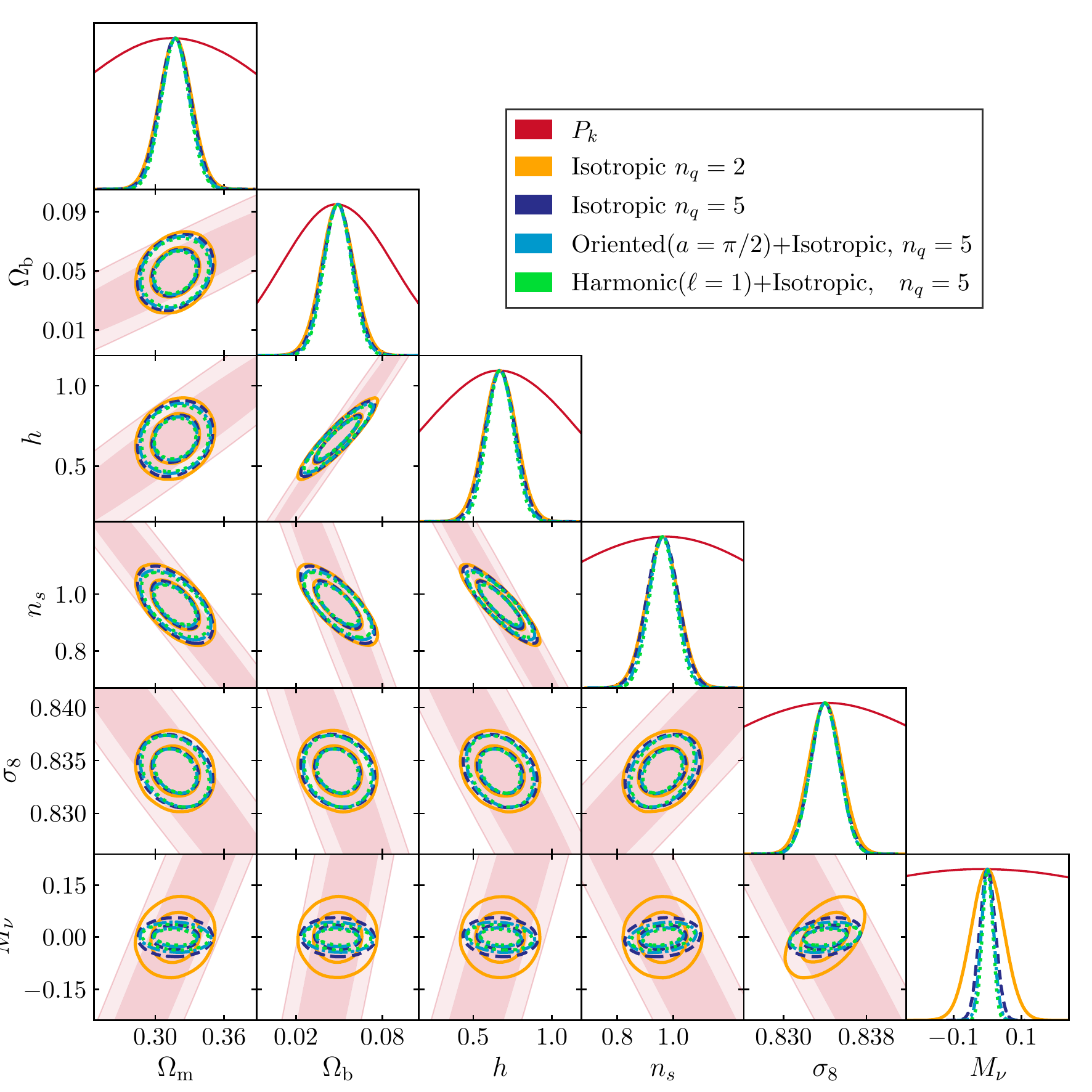}
    \includegraphics[width=.49\linewidth]{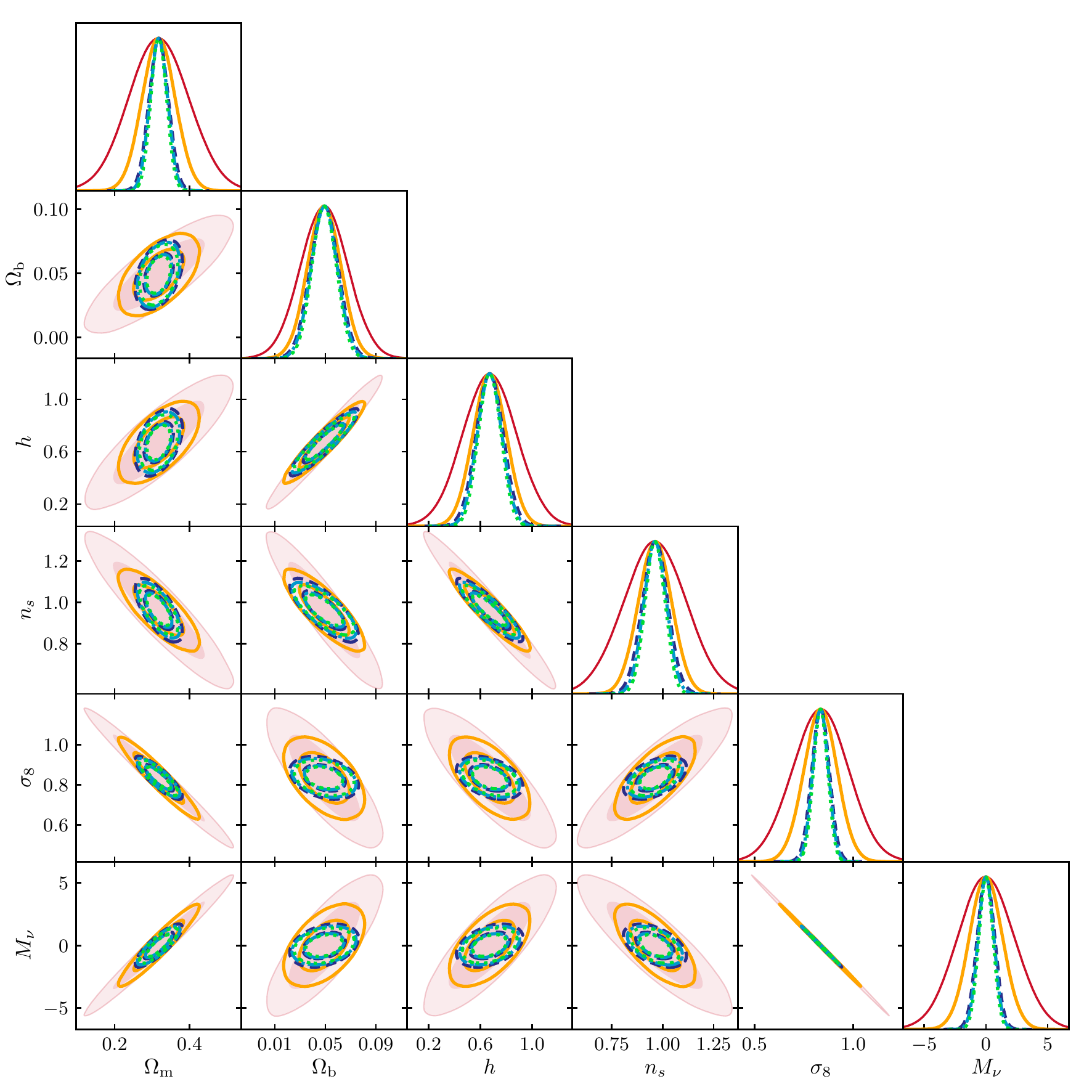}
    \caption{
     1$\sigma$ and 2$\sigma$ marginalized constraints on cosmological parameters from total matter field on the left, and cb field on the right. 
     Different lines correspond to $P_k$ (red), isotropic wavelet moduli with powers $q = 1, 2$ (orange) and powers $q = 1/4, 1/2, 1, 2, 4$ (dark blue), isotropic + oriented (light blue), as well as harmonic (green)..
    In both panels we set $k_{\rm max} = 0.5 \ h^{-1}{\rm Mpc}$.
    In both lower triangles we observe that the proposed wavelet modulus integrals achieve a strictly better constraining power than $P_k$ for all parameter pairs. They also show different correlation structure for several variable pairs.
    All statistics evaluated here are roughly set up to be extensions or subsets of each other. This is reflected in the monotonic increase in constraining power observed. Adding more integral powers is shown to have a strong effect on tightening the neutrino mass constraint $M_\nu$: Only using powers $q = 1$ and $q = 2$ shows a correlation between $M_\nu$ and $\sigma_8$ that is stronger than $M_\nu$ with any other variable. Adding more powers decreases this correlation. Adding angular information in the form of oriented or harmonic wavelets strengthens the constraints further, but mostly retains the correlation structure.
     Note that the strong neutrino mass constraints from the total matter field (left) are not replicated in the cb field (right), and we observe a strong correlation between the constraints for $\sigma_8$ and $M_\nu$.}
    \label{fig:ellipses}
\end{figure*}

To highlight degeneracies between cosmological parameters and the further illustrate the information content of various wavelet statistics vs. the power spectrum, we show the 2D marginalized constraints on cosmological parameters in Fig. \ref{fig:ellipses} for a subset of the results discussed above. The left panel are the constraints from the total matter field, while the one on the right shows those from cb field. Different lines correspond to isotropic wavelets with $Q=4, \ n_q=2$ in orange, isotropic wavelets with $Q=4, \ n_q=5$ in dark blue, oriented wavelets with $Q=4, \ n_q=11, \ a= \pi/2$ in light blue, and finally harmonic wavelets $l=1$ in green.
The left-hand panel shows constraints using the total matter (m) field derivatives for the neutrino mass, whereas the right-hand panel shows the constraints obtained using only cold dark matter and baryons (cb).

Notably, all constraints, and especially the neutrino mass constraints are worse on the right-hand side, and there is a strong degeneracy, manifested by a strong correlation between $\sigma_8$ and $M_\nu$, which is not significantly alleviated by the wavelet modulus integrals. 

In general, the $cb$-field constraints are worse than the $m$-field constraints, but are improved similarly by progressively including wavelet moments.

\subsection{Constraining power as a function of $k_\textrm{max}$}

We evaluate the constraining power of the proposed methods as a function of $k_\textrm{max}$, the maximum frequency permitted to be extracted from the signal, both by $P_k$ and the wavelet transform from which the moments are computed. Since the Quijote simulations are expected to be physically accurate only above a certain scale ($k\leq 0.5~h/{\rm Mpc}$), this restriction allows control over which scales are being used. 

\begin{figure*}[t]
    \centering
    \includegraphics[width=.99\linewidth]{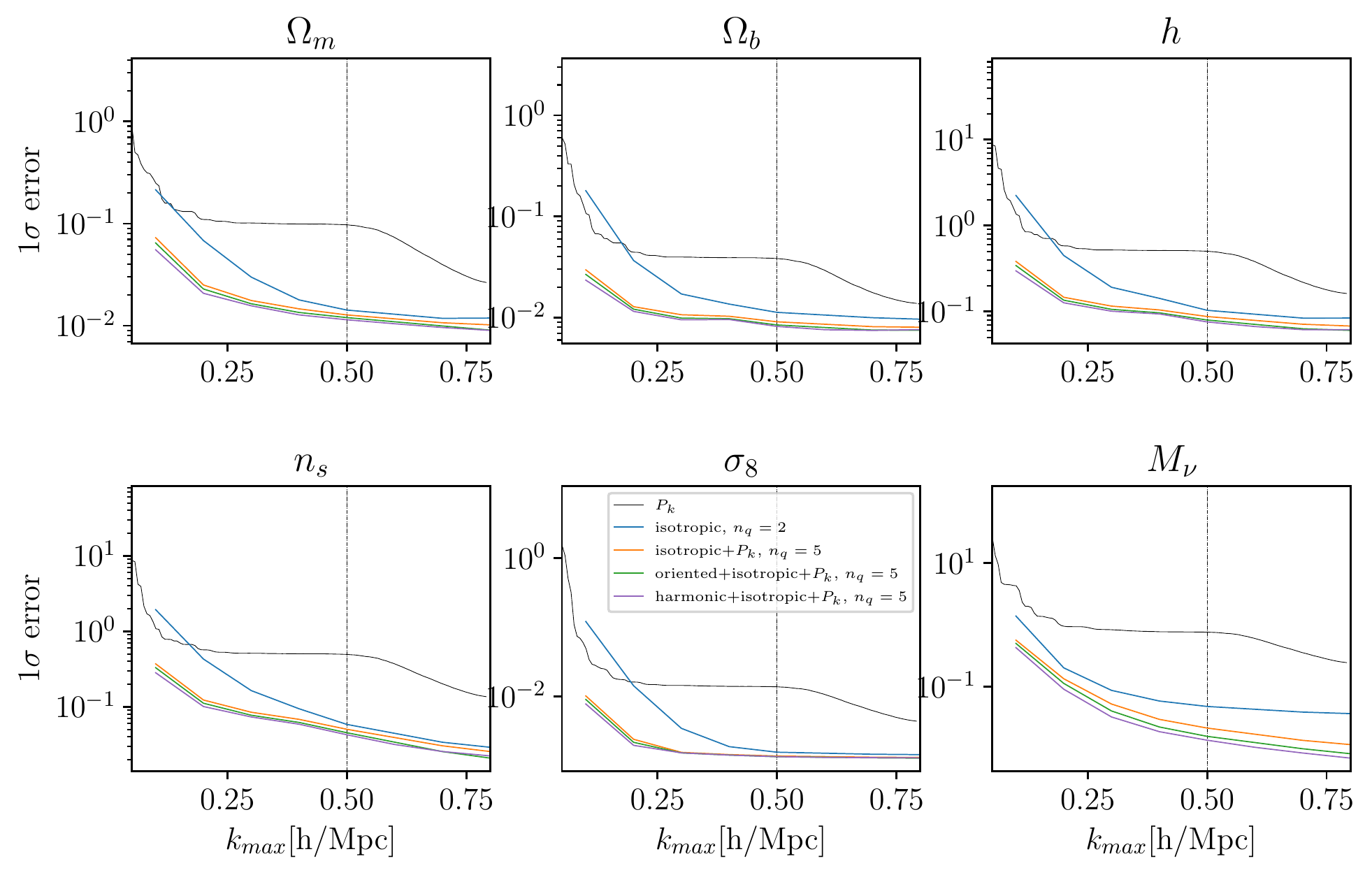}
    \caption{$1\sigma$ marginalized constraints on cosmological parameters as a function of the maximally permitted spatial frequency. 
    The cutoff frequency of $k_\textrm{max}=0.5h/\textrm{Mpc}$ is traced as a vertical line. 
    The isotropic wavelet modulus integrals with powers 1 and 2 are in blue. Starting at around $k_\textrm{max}=0.2h/\textrm{Mpc}$, their constraining power becomes significantly stronger than that of $P_k$. This is owed to the integration power 1, since integration power 2 consists of a linear combination of the $P_k$ bins. In orange, the isotropic wavelet modulus integrals with powers $1/4, 1/2, 1, 2, 4$ are shown to constrain stronger than using only powers $1,2$, for every value of $k_\textrm{max}$. The difference at the $k_\textrm{max}$ line is strongest for $M_\nu$. Including an invariant wavelet modulus descriptor based on oriented wavelets (green) provides another increment in constraining power, across the board, with the most significant improvement again in neutrino mass $M_\nu$. Using harmonic wavelets instead of oriented wavelets (purple) provides a slightly better constraint. 
    }
    \label{fig:kmax}
\end{figure*}

In Fig.~\ref{fig:kmax} the evolution of the constraints for the six parameters is shown for the power spectrum and the same three selected statistics as above. We first observe that every graph is monotonically decreasing -- the higher the allowed maximal frequency, the better the constraint. This is necessarily the case for the power spectrum, because bins are added, but less so for the wavelet statistics, because their number remains constant, since we simply scale the wavelet pyramid to adjust to the smallest scale (non-informative, ``bunched-up" larger scales can be removed a posteriori).

Next, we observe that for almost all values of $k_\textrm{max}$, and for all values of $k_\textrm{max}$ greater than $0.2~h/{\rm Mpc}$%
, the proposed methods provide vastly better constraints than $P_k$ alone. For $M_\nu$ the various proposed methods differ the most -- the inclusion of more integral powers and wavelet types making a significant difference here.
The vertical line denotes the cutoff of $k_\textrm{max} = 0.5~h/{\rm Mpc}$, the band limit below which physical plausibility is ensured.
(Note the improvement even $P_k$ achieves beyond this limit.)

\subsection{Derivatives of wavelet features}

\begin{figure}
    \centering
    \includegraphics[width=.99\linewidth]{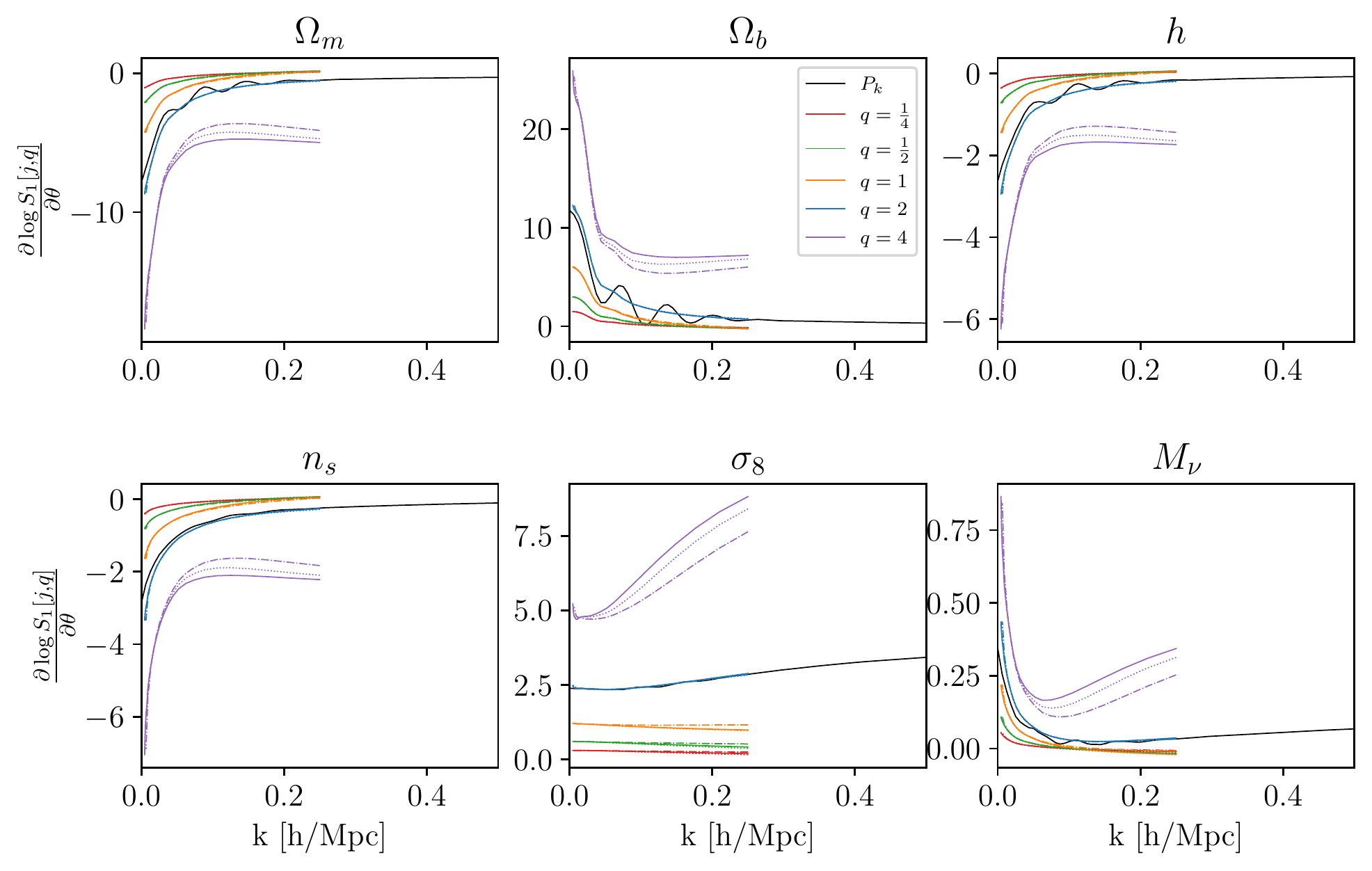}
    \caption{Visualization of the partial derivatives of the logarithms of the statistics at the fiducial cosmology. 
    All the wavelet integrals are indexed by their frequency center on the k-axis. High-frequency wavelets reach $k=0.5 \ h/{\rm Mpc}$, but also k=0, averaging at $k=0.25 \ h/{\rm Mpc}$.
    Most statistics show a decrease in derivative magnitude with increasing frequency. Power spectrum is shown in black and exhibits slight to medium-sized oscillations. 
     The curve for integral power 2 closely follows the power spectrum, without the oscillations due to smoothing. Integral powers lower than 1 tend to have lower magnitudes and integral power 4 has a higher magnitude than power spectrum. In the case of $\sigma_8$ and $M_\nu$, the low-power derivatives 1/4, 1/2 and 1 show opposite tendencies than powers 2 and 4 with increasing k.  Dash-dotted lines show the derivatives of an oriented wavelet (angular width $\pi/8$), dotted lines show the derivatives of a harmonic wavelet ($l=4$). The different statistics show a diverse set of behaviors, likely contributing to disentangling effects of different parameters.
    }
    \label{fig:derivatives}
\end{figure}

To further investigate the sensitivity of each summary statistic to individual cosmological parameters and the dependence of the constraint on the choice of $k_{\rm max}$, we show the log-derivative of $P_k$ and a selection of wavelet moments in Fig. \ref{fig:derivatives}. 

Power spectrum (black) plots globally decay in magnitude towards zero, but exhibit baryon acoustic oscillations (BAO) with a period of around $0.1~h/{\rm Mpc}$. 

Wavelet moments are shown for different powers (differentiated by color) and wavelet type (isotropic: solid, oriented: dash-dotted, harmonic: dotted). 
Values of $k$ correspond to the wavelet frequency centers. Recall that the radial bump function is symmetric around the frequency center. Hence, the maximum-frequency wavelet, which touches the cutoff at $k_\textrm{max}=0.5 \ h/\textrm{Mpc}$, has its frequency center at $k = 0.25 \ h/\textrm{Mpc}$, and is plotted at this location.

We first observe that for most powers, wavelet type does not distinguish the derivative plots very much. For power $q=4$ (purple), the different wavelet types exhibit the strongest differences amongst each other. We further note that wavelet moments at power $q=2$ (blue) trace a smoothed version of the power spectrum line (black). 
This is due to the fact that the isotropic wavelets at power $q=2$ are linear combinations of the power spectrum coefficients. 
Since the linear combinations span across wide bands of power spectrum, this also explains the smoothing of the BAO features.
Observe next that the derivatives of small integral-power coefficients have smaller magnitudes, whereas the large integral-power coefficients have larger-magnitude derivatives. The shape of the derivatives with respect to $k$ are generally similar, with the notable exceptions of $M_\nu$ and $\sigma_8$, where the low-power derivatives always slope downward, whereas power spectrum and $q = 4$ slope upward with higher $k$. 
This difference in pattern may be linked to the constraint improvement with low powers.

\begin{figure}
    \centering
    \includegraphics[width=.99\linewidth]{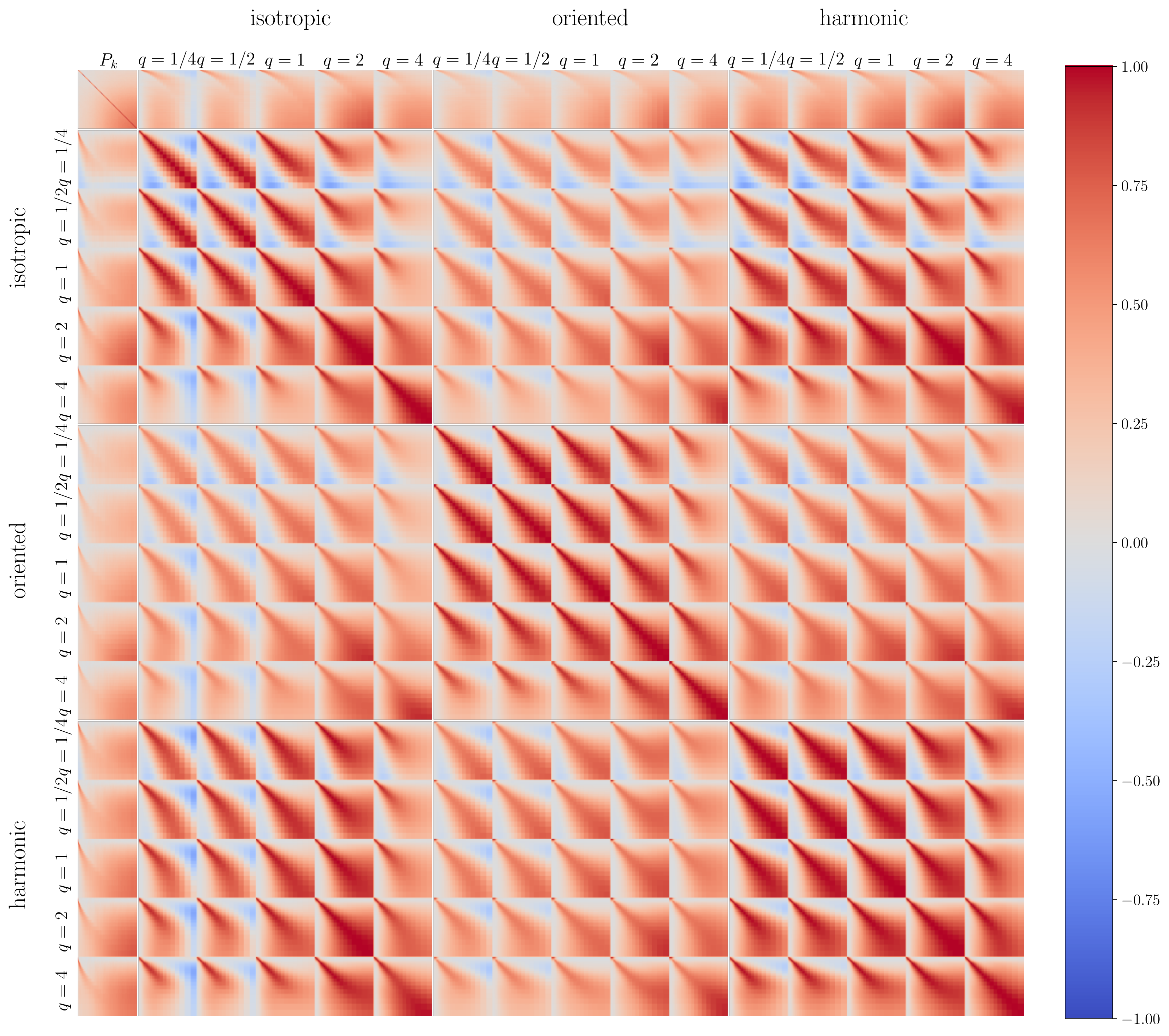}
    \caption{
    Visualization of the correlation matrix of power spectrum, 5 different integral powers of isotropic wavelet, oriented wavelet (angular width $\pi / 8$), and harmonic wavelet ($l=4$). 
    The top left corner shows the power spectrum correlation. The next five blocks show the isotropic wavelet, followed by oriented and harmonic wavelets. 
    Note how low powers create an anticorrelation between high-frequency and low-frequency wavelets. Note also the absence of cross-correlation between high frequencies of low and high integral powers. These signs of decorrelation indicate the the information provided by these features is complementary.}
    \label{fig:covariance}
\end{figure}

In Fig. \ref{fig:covariance}
the correlation matrix 
for several groups of coefficients is visualized. The groups are power spectrum, isotropic wavelet modulus integral, oriented wavelet modulus integral, harmonic wavelet modulus integral. They are separated by white lines.
Apart from the power spectrum, each of the wavelet moments is shown for five integral powers ($q = 1/4, 1/2, 1, 2, 4$).
In order to mimic the linear spacing of the power spectrum bins, the wavelet moments are sized according to the corresponding wavelet width. 
This means that, owing to their larger frequency span (see Fig. \ref{fig:wavelets}), the high-frequency wavelets occupy proportionally wider squares to represent the range of frequency content they extract. On the other hand, the low-frequency wavelets use the equivalent of fewer $P_k$ bins and are depicted on smaller squares.

We observe in the top row that $q = 2$ is most correlated with the power spectrum and in general observe that the power spectrum correlates more with low frequencies wavelets than with high ones.
The correlation pattern between powers is generally repetitive, but some differences are visible.
For example, small integral powers lead to an anticorrelation between small and large scales. This is most visible in the isotropic wavelet modulus integrals.
Further, we observe that the higher the frequency, the less correlated are the low powers ($q = 1$ or lower) with the high powers ($q = 2$ or higher).
This is likely another manifestation of relevant, uncorrelated features being extracted by the lower powers.

\subsection{Reducing the number of features}\label{sec:feature_reduction}
\begin{figure}
    \centering
    \includegraphics[width=0.99\linewidth]{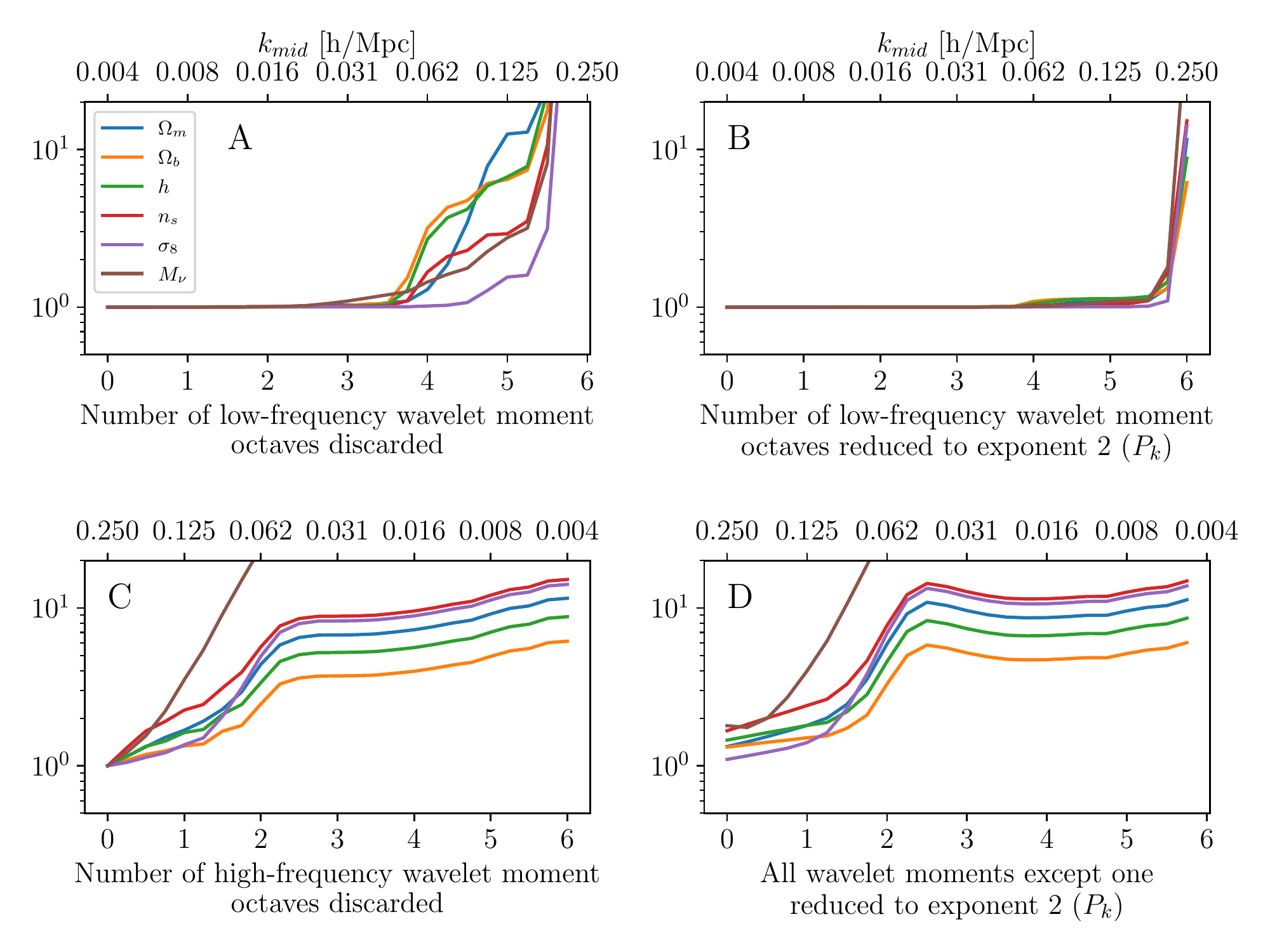}
    \caption{Visualization of feature reductions. In every plot, the degradation of constraining power is plotted relatively to the one with the full set of coefficients, i.e. the value of 1 indicates no degradation, any higher value indicates a degradation by that factor. Panel A: Wavelet scales are progressively removed starting from the largest scale. Observe that two full octaves can be removed without degradation; Panel B: Progressive removal of wavelet moment exponents scale by scale, starting from largest, leaving only exponent 2. This retains power spectrum information. Observe a first degradation after removal of 3-4 octaves ($k_{mid} \approx 0.06$), and a drastic degradation when using only exponent 2 (rightmost value); Panel C: Progressive removal of small-scale/high-frequency wavelet moment exponents, leaving only exponent 2. Observe the immediate and continuing degradation of constraints when reducing any of the smaller scales; Panel D: Reduce to exponent 2 all scales but one, indexed on the x-axis. Retaining other exponents for smallest scales is best. Note that the transition is gradual, not abrupt and only in highest frequency.}
    \label{fig:coefficient_reduction}
\end{figure}

The results on Gaussianity of the filtered fields exhibited in Fig.~\ref{fig:quijote_vs_gaussian} show clearly that the smallest two scale octaves are substantially non-Gaussian in distribution, whereas on larger scales the wavelet moments behave like those of a Gaussian field. As explained in section \ref{sec:analytic_properties}, Gaussianity of a field in particular means that the different wavelet moments are very simple functions of each other, and hence are redundant with each other. This means that for large, Gaussian-like scales, it should be possible to remove all exponents except one, and retain all constraining information. We performed this feature reduction analysis, removing scale by scale, starting from the coarsest, and retaining only exponent 2, which amounts to retaining power spectrum information. Results are shown in Fig.~\ref{fig:coefficient_reduction}, panel B, where we show all constraints relatively to their values with full coefficients. We observe that all constraint curves are flat equal to 1 up to the last two scale octaves ($0.06~h/{\rm Mpc}$), implying that power-spectral information is sufficient at these scales. The degradation on the finest two octaves shows that on these scales, the exponents different from 2 add important information to the constraints.
Panel A shows a full removal of the largest scales (not even retaining power spectrum information). Here it can be seen that the largest two octaves can simply be removed without affecting the constraints in any way. From this we conclude that $J=4$ is a good choice for analyzing Quijote $256^3$ voxel cubes with $k_{max}=0.5~h/{\rm Mpc}$.
Panel C performs the same reduction as panel B (retaining exponent 2), but starting with removal of small scales. It can be seen that degradations happen immediately upon removal of the smallest scales and continue accumulating the more scales are removed.
Panel D keeps only exponent 2 for all scales, and adds other exponents for only one scale. It can be seen that adding exponents to the smallest scales leads to the strongest constraint improvement. However, it should be noted that the improvement is not an artifact of the smallest scale (as one might read from panel B), but rather provides improvement in a continuous manner across scales.

\begin{figure}[t]
    \centering
    \includegraphics[width=.99\linewidth]{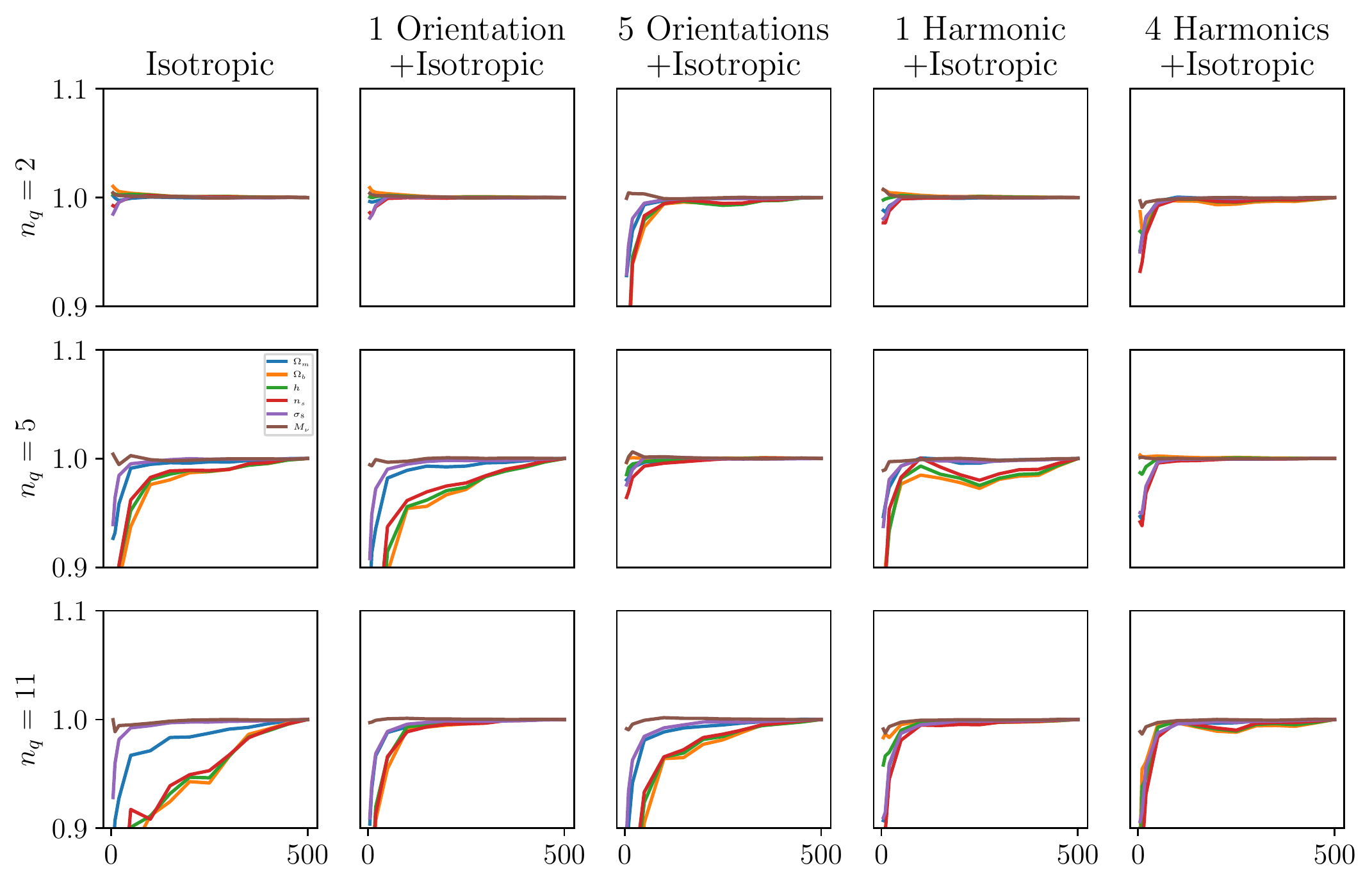}
    \caption{Stability of the empirical Cram\'er-Rao bound estimate as a function of sample size for the derivative estimator. Stability of the CR estimate reassures us that enough data points have been used and the estimator can be considered converged. 
    Every plot shows the CR bound for every parameter as a fraction of its value at full data use, as a function of number of data points used for the estimate. 
    Columns show different combinations of wavelet types. Rows show different collections of integral exponents.
    Globally, a reduction by half the amount of points (500 to 250) results in much less than 10\% change for any of the feature sets evaluated. However, we observe that some plots show stronger deviation from the reference value than others. Generally, the more integral powers are used, the less stable the estimate. The more wavelet types used, then more stable.}
    \label{fig:stability}
\end{figure}

\section{Conclusions}

In this paper we have introduced band-limited and polar-separable 3D wavelets, and used them to compute wavelet 
moments.
We investigate some of their properties and assess their information content in constraining cosmological parameters using the Quijote data set, a large suite of dark matter N-body simulations.
Wavelet moments are summary statistics that can be extracted from texture-like stationary signals, such as cosmological matter density fields. They are created by integrating the modulus of the wavelet-transformed signal, raised to a power.
These descriptors have their origin in the multifractal literature and applications to fluid dynamic turbulence.
Apart from constraining cosmological parameter values, we use the property that wavelet moments are log-linear in the exponent up to an additive function to show that the small scales of the Quijote fields are not Gaussian.
The polar-separable 3D wavelets were presented in three flavors, isotropic, oriented, and harmonic.
In progressive experiments it was shown that isotropic wavelets achieve a large improvement (factor 5-10) compared to power spectrum when integral powers other than $q = 2$ were added -- this was the strongest effect.
Further improvement was achieved by incorporating orientation information.

This approach has a commonality with the marked power spectrum in that they both investigate fields modified through exponentiation.
Indeed, using integral powers of below $q = 1$ acts as a compression: peaks are lowered, and non-zero small values are raised closer to one. In the zero-power-limit it would lead to a binary mask showing where the filtered field is zero and nonzero. Integrating over it would simply count the nonzero voxels. Using integral powers above $q = 1$ highlights peaks and suppresses small values. Again, in the infinite limit, the integrals would essentially be measuring the size of the set of peaks of the filtered fields. It is possible that filtering and the different powers used could approximate the values of the marked power spectrum. 

However, there are also important differences between wavelet modulus integrals and the marked spectrum: 
First, the multiplicative operation marking the field, since it becomes a convolution in Fourier space, may lead to mixing of higher-frequency information into the lower frequency ranges. 
Second, computing the Fourier power spectrum with linear bins suffers the same instabilities as the non-marked power spectrum under small deformations, leading to higher variance in high frequencies.

While in this work we emphasize the importance of integral powers, future work may explore the additional benefit of second-order wavelet-based features, such as scattering transforms and wavelet phase harmonics, all in the 3D setting.
As this work was being completed, a paper by \cite{valogiannis2021optimal} appeared on arXiv, in which they use solid harmonic wavelet scattering \citep{eickenberg_solid_2018-1} to obtain similar parameter constraints on the Quijote suite. Because solid harmonic wavelets are based on Gaussians, their support both in the Fourier domain and in the signal domain is infinite. Therefore, it is difficult to compare the obtained constraints to the ones obtained here, under strict band limitation. In this recent work, while the $\ell=0$ harmonic (a Gaussian) is largely contained below the cutoff rate of $k_\textrm{max}=0.5~h/{\rm Mpc}$, the higher-harmonic-order wavelets with $\ell > 0$ have much wider supports in frequency space given the same Gaussian width, and may use frequency information up to, and beyond $k_\textrm{max}=0.8~h/{\rm Mpc}$. If we adjust our band-limited wavelets to explore these higher bands, we obtain similar constraints. If we restrict them to $k_\textrm{max}=0.5~h/{\rm Mpc}$, then our neutrino mass constraint values lie at around 150\% of the ones reported in \cite{valogiannis2021optimal}.

Furthermore, all these methods can be extended to work in redshift space by decoupling the redshift axis from the other two. We expect  improvements in constraints from light cones, especially for the $cb$ field \cite{Adrian_fake_nu}. Beyond this, extensions to be able to analyze galaxy clustering from upcoming surveys are possible. In these cases, simulation-based inference will likely be required, since one will not have access to the derivatives around fiducial cosmology to compute Fisher information.

\label{sec:conclusions}

\section*{Acknowledgments}
A.M.D. is supported by the Tomalla Foundation for Gravity and the SNSF project ``The  Non-Gaussian  Universe and  Cosmological Symmetries", project number:200020-178787. 
P.L. acknowledges STFC Consolidated Grant ST/T000473/1.

\bibliography{bibtex}{}
\bibliographystyle{aasjournal}

\appendix

\section{Reconditioning of the coefficient spaces}\label{sec:reconditioning}

While mathematically the Fisher information is invariant to invertible affine transformations, in numerical practice it may not be. For example, if extracted features are highly correlated, or differ in amplitude by several orders of magnitude, the empirical covariance matrix may be ill-conditioned. To avoid this, we make use of the affine invariance to precondition the features using whitening based on principal component analysis (PCA). For this we use the mean, standard deviation, and covariance extracted from the fiducial simulations.

Concretely, we adopt the following procedure. Let $X_0\in\mathbb R^{n_\textrm{fiducial}\times n_\textrm{descriptors}}$ be the set of summary statistics for the fiducial fields, and $X_d\in\mathbb R^{n_\textrm{derivatives}\times n_\textrm{descriptors}}$ any derivative statistic. We determine the mean across the fiducial simulations for every coefficient by computing $\bar X_{0, j} = \sum_i X_{0, ij} / n_\textrm{fiducial}$. We subtract this mean value from all the coefficients (fiducial $X_0 - \bar X_0$ and derivatives $X_d - \bar X_0$). This change leaves all derivatives and the covariance matrix invariant. Next, we determine the standard deviation of every coefficient $s_{0, j}^2=\sum_i (X_{0,ij} - \bar X_{0,j})^2 / (n_\textrm{fiducial} - 1)$, again on the fiducial set. We divide every coefficient (from the fiducial and derivative data sets) by the corresponding standard deviation, leading to $(X_0 - \bar X_0) / \sqrt{s_0^2}$ and $(X_d - \bar X_0) / \sqrt{s_0^2}$. This amounts to a diagonal conditioning of the covariance matrix and leaves the Fisher matrix invariant. Lastly, we apply an orthogonal transformation to all the features, which consists of the eigenvectors of the fiducial covariance matrix. If $USV^T = (X_0 - \bar X_0) / \sqrt{s_0^2}$ is the SVD of the standardized fiducial descriptors, then we compute the desired transformation as $\left((X_0 - \bar X_0) / \sqrt{s_0^2}\right)V$. This diagonalizes the covariance matrix and leaves the Fisher information invariant.

We further elect to remove feature dimensions with the smallest variance until the conditioning number of the remaining matrix (the square-root of the quotient of the largest and smallest non-zero eigenvalue) is no bigger than $10^7$. This corresponds to removing columns from $V$ that correspond to low eigenvalues. Since this effectively projects our feature onto a smaller-dimensional subspace, the resulting Fisher matrix constraints are still valid, if slightly worse compared to those obtained from the full feature set.

\end{document}